# Pasture Intake Protects Against Commercial Diet-induced Lipopolysaccharide Production Facilitated by Gut Microbiota through Activating Intestinal Alkaline Phosphatase Enzyme in Meat Geese

Qasim *Ali*[1], Sen *Ma*[1,2,3], Umar *Farooq*[4], Jiakuan *Niu*[1], Fen *Li*[1], Muhammad *Abaidullah*[1], Boshuai Liu[1], Shaokai *La*[1], Defeng *Li*[1,2,3], Zhichang *Wang*[1,2,3], Hao *Sun*[1,2,3], Yalei *Cui*[1,2,3], and Yinghua *Shi*[1,2,3,*]

[1] Department of Animal Nutrition and Feed Science, College of Animal Science and Technology, Henan Agricultural University, Zhengzhou, Henan, 450002, China

[2] Henan Key Laboratory of Innovation and Utilization of Grassland Resources, Zhengzhou, Henan, 450002, China

[3] Henan Herbage Engineering Technology Research Center, Zhengzhou, Henan, 450002, China

[4] Department of Poultry Science, University of Agriculture Faisalabad, Sub Campus Toba Tek Singh, 36050, Pakistan

**\*Correspondence:**

 Yinghua Shi

annysyh@henau.edu.cn

## Abstract

Diet strongly affects gut microbiota composition, and gut bacteria can influence the cecal barrier functions and systemic inflammation through metabolic endotoxemia. In-house feeding system (IHF, a low dietary fiber source) may cause altered cecal microbiota composition and inflammatory responses in meat geese via increased endotoxemia (lipopolysaccharides) with reduced intestinal alkaline phosphatase (ALP) production. The effects of artificial pasture grazing system (AGF, a high dietary fiber source) on modulating gut microbiota architecture and gut barrier functions have not been investigated in meat geese. The intestinal ALP functions to regulate gut microbial homeostasis and barrier function appears to inhibit pro-inflammatory cytokines by reducing LPS-induced reactive oxygen species (ROS) production. The purpose of our study was to investigate whether this enzyme could play a critical role in attenuating ROS generation and then ROS facilitated NF-κB pathway-induced systemic inflammation in meat geese. First, we assessed the impacts of IHF and AGF on gut microbial composition via 16 sRNA sequencing in meat geese. In the gut microbiota analysis, meat geese supplemented with pasture demonstrated a significant reduction in microbial richness and diversity compared to IHF meat geese demonstrating antimicrobial, antioxidation, and anti-inflammatory ability of AGF system. Second host markers analysis through protein expression of serum and cecal tissues and quantitative PCR of cecal tissues were evaluated. We confirmed a significant increase in intestinal ALP-induced Nrf2 signaling pathway representing LPS dephosphorylation mediated TLR4/MyD88 induced ROS reduction mechanisms in AGF meat geese. Further, the correlation analysis of top 44 host markers with gut microbiota shows that artificial pasture intake induced gut barrier functions via reducing ROS-mediated NF-κB pathway-induced gut permeability, systemic inflammation, and aging phenotypes. In conclusion, AGF system may represent a novel therapy to counteract the chronic inflammatory state leading to low dietary fiber-related diseases in animals.

## 1. Introduction

Intestinal homeostasis seems to be a defining factor for poultry health that is affected by oxidative stress either produced by heat stress or feed stress [1]. The poultry birds such as broilers, layers, geese, and turkeys are continuously exposed to lipopolysaccharides (LPS) via different routes such as feed, water, and fine dust particles in the house that always contain some amounts of LPS. However, the major natural source of LPS is the complex community of gram-negative bacteria in the intestines [2]. LPS is an outer membrane component of gram-negative bacteria such as *Enterobacteriaceae* [3,4], *Escherichia coli* (*E. coli*) [5], *bacterodales* [6], and *cyanobacteria* [7] that are recognized by toll-like receptors (TLRs), particularly TLR4 and then invade the intestinal tissues and get access to the bloodstream thereby provoking reactive oxygen species (ROS)-induced systemic diseases [8-11]. This leads to a leaky gut [12] that causes diarrhea, decreased nutrient absorption, and internal fluid loss in broilers [13,14].

From different studies, two supporting shreds of evidence suggest that nuclear factor kappa B (NF-κB) is activated either by ROS [15] or LPS [16]. Further, LPS [17,18] and ROS [19] phosphorylate NF-κB inhibitor α (*IκB-α*) and let the NF-κB migrate to the nucleus and then exert their inflammatory and apoptotic impacts through activating NF-κB pathway [20,21]. When NF-κB pathway is established, then it consistently contributes to inducing chronic low-grade inflammation [22-24] through upregulating oxidative stress [25], pro-inflammatory mediators inducible nitric oxide (*iNOS*) and cyclooxygenase (*COX*)-2 [26,27] and pro-inflammatory cytokines (*IL-1β*, *IL-6*, and *TNF-α*) [28-30].

Gut microbial-induced LPS-mediated NF-κB activity orchestrates chronic low-grade inflammation that is a major disease risk factor in today's animals' life [31,32]. Thus, modifiable factors that can reduce LPS-induced inflammation may potentially modulate disease risk. Different possible modifiable factors such as feed antibiotics [33], dietary fiber and threonine [34], pectin [35], Schisandra A [23], phytochemicals [36], and oral alkaline phosphatase [37,38] have been used in animals to detoxify LPS. Feed antibiotics increase food-born pathogenic bacterial-induced LPS resistance [39,40] while phytochemicals used as anticancer drugs show minimum side effects in animals [36]. Albeit different dietary fibers detoxify LPS-induced chronic low-grade inflammation [34,35], nobody did clearly explain the mechanisms of how dietary fibers carry out these processes.

It has been known that different dietary fiber sources such as glucomannan, oligosaccharides, sialyl lactose, and galactooligosaccharides shape the gut microbiota to regulate intestinal ALP in rats and other animals [41-44]. Intestinal ALP is an endogenous antimicrobial peptide that is secreted from the apical enterocytes of small intestine and then moves towards the large intestine [45]. Intestinal ALP has been shown to dephosphorylate LPS, CpG-DNA, and flagellin [39,46]. In previous studies, lipid A moiety plays a role of bridge in activating myeloid differentiation factor 88 (MyD88) pathway by binding LPS with TLR4 and then activating oxidative stress [3,47] and NF-κB-induced systemic inflammation [22,23]. To what extent and in which way the intestinal ALP dephosphorylates LPS by breaking TLR4/MyD88-induced ROS production and NF-κB-induced systemic inflammation is not defined clearly in previous studies. Furthermore, intestinal ALP promotes healthy homeostasis of gut microbiota [37,48,49] and gut barrier functions [28,38,50], which has been associated with lowering systemic inflammation in several studies of healthy adults [51] and their specific health conditions (e.g. obesity [52], diabetes [53], and cardiovascular disease [54]). The increasing evidence of the salutary functions of ALP underlines the significance of the naturally occurring brush border enzyme. Therefore, there is a need to find optional nutritional strategies that could naturally induce and regulate the endogenous growth of intestinal ALP in poultry birds. One of the optional possible nutritional strategies is the application of dietary fiber that has been used in different studies to induce intestinal ALP production [41,42,44].

Geese are herbivorous, and because of their unique ability to use high fiber feeds, pasture was suggested to be included to promote health [55]. In China, the most commonly available pasture is ryegrass which is rich in protein, dietary fiber, fatty acids, iron, zinc, magnesium, calcium, vitamins, essential amino acids, alkaloids, steroids, flavonoids, glycosides, phenols, and tannins [56,57]. Recently, several reports suggest that ryegrass could regulate intestinal microflora of Beijing-you chickens [58] and improve ethnomedical properties like being antioxidant, antimicrobial, and anti-inflammatory diseases in animals [59,60]. However, to our knowledge, formal mediation analysis testing whether and to what extent inflammation mediates the observed inverse association between high dietary fiber intake and LPS-induced oxidative stress is lacking. Moreover, the kelch-like ECH-Associating protein 1-nuclear factor erythroid 2 (Keap1-Nrf2) system plays a central role in the regulation of the oxidative stress response, and that NRF2 coordinately regulates cytoprotective genes [61]. Here we reported that the regulation of gut microbial-induced endogenous intestinal ALP by pasture intake preserves the normal homeostasis of gut microbiota, dephosphorylates LPS/TLR4/MyD88 pathway-mediated ROS insults, and activates KEAP1-NRF2 pathway to deteriorate NF-κB-induced systemic inflammation in meat geese.

## 2. Materials and Methods

### 2.1. Animals, Diets, and Housing

A total of 180, 25-day-old Wanfu mixed-sex geese from the commercial geese farm were purchased from Henan Daidai goose Agriculture and Animal husbandry development Co. LTD (Zhumadian, China). The geese with an average weight of 693.6 g ($\pm$3.32) were divided into two homogeneous groups: (1) in-house feeding group (IHF, $n = 90$) and (2) artificial pasture grazing group (AGF, $n = 90$), 12 h artificial pasture grazing (06:00-18:00 h) with once a day (19:00 h) in-house feeding group. Each group consisted of six replicates with 15 geese per replicate. All the geese had free access to feed and fresh water *ad-libitum*. The IHF group meat geese were fed a commercial diet (**Table 1**). Two diets were used: a grower diet (25 to 45 days) and a finisher diet (46 to 90 days). The artificial pasture grazing system was established in form of grazing of meat geese at the expense of ryegrass. The nutritional composition of ryegrass was DM (90%), crude protein (15.47%), ash (8%), NDF (65%), ADF (38%), EE (3.3%), calcium (0.90%), and phosphorous (0.47%). The experiment lasted for 66 days (**Supplemental Figure 1**).

All procedures were approved by the Research Bioethics Committee of the Henan Agricultural University (approval # HENAU-2021).

### 2.2. Sample Collection

Body weight and feed intake were measured every week. The pasture feed intake was measured using method described by Cartoni Mancinelli et al. [62]. On day 45, 60, and 90, we selected six healthy meat geese per replicate with a body weight range of $\pm$ 1 std. from mean 1.63-2.31 kg, 3.33-4.28 kg, and 4.38-5.39 kg respectively. Blood samples were collected in non-anticoagulant sterile blood vessels from the jugular vein. Serum samples were then obtained by centrifuging the blood samples at 4,000 × g for 15 min at 4 °C and stored at –80 °C until analysis. After blood sampling, the geese were slaughtered and pH was determined from the proventriculus, gizzard, ileum, and cecum. Fresh cecal chyme was collected from the cecum using sterile 5-ml centrifuge tubes and then stored at –80 °C for further analysis. The cecal tissues were immediately removed, thoroughly washed with phosphate-buffered saline (PBS), stored in liquid nitrogen, and then preserved at –80 °C for further analysis. Intestinal segments such as cecal tissues were fixed by immersion in 10% buffer neutral formalin.

## 2.3. Measurement of LPS, ROS, and ALP Levels

The serum and cecal tissues were sampled to measure lipopolysaccharides (LPS), reactive oxygen species (ROS), and alkaline phosphatase (ALP) activities. The kits were purchased from Shanghai Enzyme Link Biotechnology Co, Ltd (Shanghai, China) and all experimental procedures were performed according to the manufacturer's instructions.

## 2.4. Bacterial Growth Conditions

Batch cultivation for cecal *Escherichia coli* (*E. coli*) was carried out in Luria broth (LB) medium at 37 °C with a 2-L working volume. LB medium was from recipe of Miller (5 g yeast extract, 10 g peptone tryptone, 10 g NaCl) [63]. The pH was maintained at 6.95 automatically by titration with 5% H2SO4 or 5% NaOH. Ampicillin was added to control the growth of other bacteria. Peptone tryptone and yeast extract were from OXOID, NaCl from Sigma, agar, and ampicillin from Solarbio (life sciences). The medium was made in distilled water and autoclaved under standard conditions. Dissolved oxygen in the culture was maintained at 40% saturation automatically by varying the speed of impeller rotation. Culture growth (OD600) was monitored with a DU640 Spectrophotometer (Beckman). Further, *E. coli* was cultured onto the Petri dishes for 24hrs at 37 °C. The CFU/g stool for *E. coli* from the Petri dishes was counted.

## 2.5. Measurement of Gut Permeability

The gut permeability was measured by determining the tight junction proteins ZO-1, Occludin, and Claudin concentrations from the serum and cecal tissues. The kits were purchased from Shanghai Enzyme Link Biotechnology Co, Ltd (Shanghai, China) and all experimental procedures were performed according to the manufacturer's instructions. Further, mRNA expression levels of 2 genes encoding tight junction proteins *dlg1* and *E-cadherin* were also measured from cecal tissues for gut barrier functions.

## 2.6. Measurement of Antioxidant Parameters

Heme oxygenase 1 (HO-1) and glutathione reductase (GSR) were measured from serum using ELISA kits (Shanghai Meilian Biology Technology, Shanghai, China). The total superoxide dismutase (T-SOD, #A001-1), glutathione peroxidase (GSH-Px, #A005), total antioxidant capacity (T-AOC, #A015-2–1), malondialdehyde (MDA, #A003-1), and catalase (CAT, #A007-1) were measured using diagnostic kits (Nanjing Jiancheng Bioengineering Institute, Nanjing, Jiangsu, P. R. China) according to the manufacturer's instructions.

## 2.7. Measurement of Metabolic (Plasma Lipid) Profiles

Serum total cholesterol (T-CHO, #A111-1-1), low-density lipoprotein cholesterol (LDL-C, #A113-1-1)**,** high-density lipoprotein cholesterol (HDL-C, #A112-1-1), triglycerides (TG, #A110-1-1), and blood urea nitrogen (BUN, #C013-1-1) was enzymatically determined using a kit from Nanjing Jiancheng Bioengineering (Nanjing, Jiangsu, P. R. China) following the manufacturer's instructions. Fasting blood glucose level was determined by strip method followed by Sannuo biosensor C.., Ltd (Changsha, P. R. China).

## 2.8. RNA Extraction and RT-qPCR

Total RNA from the cecal tissues (about 50 to 100 mg) was extracted by the addition of 1 mL of MagZol-reagent (#R4801-02; Magen Biotechnology, Guangzhou, Guangdong, China) according to the manufacturer's instructions. The concentration and purity of the total RNA were assessed using a NanoDrop 2000 UV-vis spectrophotometer (Thermo Scientific, Wilmington, USA). Subsequently, the RNA was reverse transcribed to cDNA using the ReverTra Ace® qPCR RT Master Mix with gDNA Remover (TOYOBO, OSAKA, Japan) according to the manufacturer's instructions. The cDNA samples were amplified by real-time quantitative polymerase chain reaction with ChamQ Universal SYBR qPCR Master Mix from Vazyme Biotechnology (Nanjing, Jiangsu, P. R. China). Gene-specific primers for each gene were designed using Primer3web, version 4.1.0 (**Supplemental Table 1**). PCR was performed on the C1000 Touch PCR Thermal cycler (BIO-RAD Laboratories, Shanghai, China) using ChamQ Universal SYBR qPCR Master Mix from Vazyme Biotechnology (Nanjing, Jiangsu, P. R. China) and was as follows: 40 cycles of 95 °C for 15 s and 60 °C for 30 s. All measurements will be performed in triplicate. The messenger ribonucleic acid (mRNA) expression of target genes relative to beta-actin (*β-actin*) was calculated using $2^{-\Delta\Delta CT}$ method [64].

*2.9. DNA Extraction and PCR Amplification*

According to the manufacturer's instructions, the microbial community genomic DNA was extracted from cecal chyme samples using the E.Z.N.A.® soil DNA Kit (Omega Bio-Tek, Norcross, GA, U.S.). The DNA extract was checked on 1% agarose gel, and DNA concentration and purity were determined with NanoDrop 2000 UV-vis spectrophotometer (Thermo Scientific, Wilmington, USA). The hypervariable region V3-V4 of the bacterial 16S rRNA gene was amplified with primer pairs 338F (5'-ACTCCTACGGGAGGCAGCAG-3') and 806R (5'-GGACTACHVGGGTWTCTAAT-3') by an ABI GeneAmp® 9700 PCR thermocycler (ABI, CA, USA). The PCR amplification of the 16S rRNA gene was performed as follows: initial denaturation at 95 °C for 3 min, followed by 27 cycles of denaturing at 95 °C for 30 s, annealing at 55 °C for 30 s, and extension at 72 °C for 45 s, and single extension at 72 °C for 10 min, and end at 4 °C. The PCR mixtures contain 5 × *TransStart* FastPfu buffer 4 μL, 2.5 mM dNTPs 2 μL, forward primer (5 μM) 0.8 μL, reverse primer (5 μM) 0.8 μL, *TransStart* FastPfu DNA Polymerase 0.4 μL, template DNA 10 ng, and finally ddH$_2$O up to 20 μL. PCR reactions were performed in triplicate. The PCR product was extracted from 2% agarose gel and purified using the AxyPrep DNA Gel Extraction Kit (Axygen Biosciences, Union City, CA, USA) according to the manufacturer's instructions and quantified using Quantus™ Fluorometer (Promega, USA).

*2.10. Illumina MiSeq Sequencing*

Purified amplicons were pooled in equimolar and paired-end sequenced (2 ×300) on an Illumina MiSeq platform (Illumina, San Diego, USA) according to the standard protocols by Majorbio Bio-Pharm Technology Co. Ltd. (Shanghai, China).

*2.11. Processing of Sequencing Data*

The raw 16S rRNA gene sequencing reads were demultiplexed, quality-filtered by Trimmomatic, and merged by FLASH with the following criteria: (i) the 300 bp reads were truncated at any site receiving an average quality score of <20 over a 50 bp sliding window, and the truncated reads shorter than 50 bp were discarded, reads containing ambiguous characters were also discarded; (ii) only overlapping sequences longer than 10 bp were assembled according to their overlapped sequence. The maximum mismatch ratio of overlap region is 0.2. Reads that could not be assembled were discarded; (iii) samples were distinguished according to the barcode and primers, and the sequence direction was adjusted, exact barcode matching, 2 nucleotide mismatch in primer matching. The taxonomy of each OTU representative sequence was analyzed by RDP Classifier (http://rdp.cme.msu.edu/) against the 16S rRNA database (e.g. Silva SSU128) using a confidence threshold of 0.7. The species composition and relative abundance of each sample were counted at the phylum level, and the composition of the dominant species of different groups was visualized by the package pie chart of R (version 3.3.1) software. PICRUSt is a software package for the functional prediction of 16S amplicon sequencing results which was used to determine the COG IDs related to LPS production. Spearman rank correlation coefficient was performed to construct a correlation heatmap among the highly abundant GO terms and cecal microbiota at phylum as well as genus levels mostly relevant to LPS production. Further, to determine the effect of microbiota interacting with apparent performance, redundancy analysis (RDA) was performed at the genus level using the R language vegan packet on Spearman correlation analysis (RDA 2014).

*2.12. Statistical Analysis*

Data were expressed as mean ± SEM. Statistical analyses were performed using SPSS 20.0 software (=D3 SPSS, Inc., 2009, Chicago, IL, USA www.spss.com. Data from two groups were evaluated by unpaired two-tailed student T-Test. Significance was considered to be at $P<$ 0.05. Spearman correlation analysis of the Euclidean distance was performed using GraphPad Prism version 8.3.0., and origin 2021. To compare host markers' relationships, Pearson's correlation analysis was performed by OECloud tools (https://cloud.oebiotech.cn.).

# 3. Results

*3.1. Artificial Pasture Grazing System Modulates Gut Microbiota to Inhibit LPS Synthesis Induced by In-house Feeding System*

The metagenome predicted functions classified using clusters of orthologous genes (COG) database in phylogenetic reconstruction of unobserved states (PICRUSt) software are performed to investigate the functional differences in the gut microbiota between the two feeding meat geese groups (in-house feeding group (IHF) and artificial pasture grazing group (AGF) meat geese) at different time points 45d, 60d, and 90d. A total of 4060, 4069, 4082, 3959, 4030, and 4060 COG IDs were identified in samples IHF45, AGF45, IHF60, AGF60, IHF90, and AGF90 respectively (unpublished data). To identify which strains contribute to LPS production in the meat geese gut, we focused on the GO terms and cecal microbiota at the phylum level mainly relevant to LPS biosynthesis (**Figures 1A and Supplemental Figures 2A-F**). At 45d, in the IHF meat geese, *Firmicutes* families dominated the four while *Actinobacteiota* dominated the two main COG terms related to LPS biosynthesis. At 60d, in the IHF meat geese, *Actinobacteriota* and *Proteobacteria* families dominated the four and two main COG terms related to LPS biosynthesis respectively. At 90d, only *Bacteroidota* family dominated the three main COG terms related to LPS biosynthesis in IHF meat geese.

Overall, *Firmicutes* and *Bacteroidota* families dominate both in the IHF and AGF meat geese at 45d, 60d, and 90d, but *Firmicutes* and *Bacteroidota* individually contribute 68.45% and 32.62% of the LPS biosynthesis in the IHF meat geese at 45d and 90d respectively (**Figure 1B**). In contrast, *Actinobacteriota* family is the minor contributor to LPS biosynthesis, with an average of 8.89% and 9.90% of the total LPS biosynthesis in IHF meat geese at 45d and 60d respectively. In contrast to other bacterial families, *Proteobacteria* family with a minute quantity with an average of 2.94% contributes to the total LPS biosynthesis in IHF meat geese at 60d.

The quality and types of diet make alterations in the gut microbiota in a time-dependent manner. We next examined whether the bacterial composition of the samples would correlate with the potency of activation or inhibition of these samples (**Figure 1C**). At the genus level, we determined the Spearman correlation between the GO terms and cecal microbiota falling in different phyla mainly relevant to LPS biosynthesis and those have been individually expressed in a supporting file (**Supplemental Figures 3A-E**). We identify a strong correlation between GO functions related to LPS biosynthesis and microbial composition at the genus level and could only find a moderate-to-strong correlation between a few microbial individual genera and the stimulatory potency of individual cecal LPS production in IHF meat geese (**Figure 1C**). However, we found that the abundance of very few bacterial genera *Lactobacillus* and *Ruminococcus_torques_group* following *Firmicutes* phylum show a moderate correlation with the activation of lipid A biosynthesis and LPS transportation periplasmic protein lptA at 45d in IHF meat geese (**Figure 1C**). While only single bacterial genera *norank_f__norank_o__Gastranaerophilales* following the phylum *Cyanobacteria* show a strong correlation with the activation of LPS transportation periplasmic protein lptA at 45d in IHF meat geese (**Figure 1C**). In contrast, we found a maximum of the bacterial genera *Prevotellaceae_UCG-001*, *Bacteroides*, and *Alistipes* following *Bacteroidota* phylum were contributing to the lipid A biosynthesis acyltransferase and lipid A biosynthesis at 60d in IHF meat geese (**Figure 1C**). Similarly, *Bacteroides* and *Rikenellaceae_RC9_gut_group* following *Bacteroidota* phylum follow the same trend in activating the lipid A biosynthesis acyltransferase and lipid A biosynthesis at 90d in IHF meat geese (**Figure 1C**). Notably, *Bacteroides* dominate activating lipid A biosynthesis acyltransferase at 60d and 90d in IHF meat geese (**Figure 1C**). Our results revealed that *Bacteroidota* phylum is by far the most abundant contributor to the LPS biosynthesis functions in IHF meat geese intestinal microbiota, consistent with their high abundance relative to other Gram-negative genera in the gut.

### 3.2. Inhibitory Effects of Artificial Pasture Grazing System on In-house Feeding System-induced ROS Production via LPS/TLR4/MyD88 Pathway in Meat Geese

In previous studies, LPS-induced ROS production is diet-dependent [65,66]. The mechanisms controlling this phenomenon are not defined clearly in meat geese. Here we will try to depict the pathway for ROS synthesis via LPS/TLR4/MyD88 pathway. To investigate the effects of microbial-induced LPS on ROS expression, meat geese were reared in the in-house feeding system throughout their whole lifespan and were compared with meat geese accomplished by artificial pasture grazing system. It is known that intestinal ALP is highly pH sensitive and is reduced in acidic environments like the stomach. It exhibits its biological activity more than pH 6 [67]. We observed that the proventriculus, gizzard, ileum, and cecal pH in the AGF meat geese was significantly increased compared with IHF meat geese (**Supplemental Figures 4A-D and Supplemental Table 2**). Then we measured the protein levels of intestinal ALP activity and expression levels of intestinal alkaline phosphatase gene (*ALPi*) and 2 separate alkaline phosphatase genes (*CG5150* and *CG10827*) from the meat geese. The serum ALP activity by enzyme-linked immunosorbent *assay* (ELISA) kit (**Figure 2A**) and mRNA expression of *ALPi* (**Figure 2B**) and alkaline phosphatase genes (*CG5150* and *CG10827*) (**Figure 2C**) increased significantly in AGF meat geese as compared to IHF meat geese at 45d, 60d, and 90d. Furthermore, intestinal ALP may contribute to maintaining the normal gut microbial homeostasis by suppressing the *E. coli* and as well as detoxifying the LPS [68,69]. To identify, whether *E. coli* contributes to activating the LPS and the suppression of intestinal ALP, we cultured the *E. coli* onto the Petri dishes for 24hrs at $37^0$C. For this, we counted the colony-forming units (CFU) for *E. coli* from the Petri dishes. We observed that the CFU/g stool was less in the AGF meat geese compared with the IHF meat geese (**Supplemental Figure 5A**). To further verify our results, we pick up one colony from the petri dish and incubate it in the Luria-Bertani (LB) medium for 48hrs at $37^0C$. Then we determined the *E. coli* cell cultures based on spectrophotometer readings at OD600 for 48hrs with an interval of 2hrs. We found a significant decline in the concentration of *E. coli* cell cultures from the AGF meat geese as compared to IHF meat geese (**Supplemental Figure 5B**).

Impairment of intestinal ALP plays a crucial role in activating lipid A moiety so that it permits LPS to bind with *TLR4* and then activate MyD88 dependent pathway [3,47]. In fact, LPS could bind to LPS binding protein (LBP), and LBP-LPS complex is transferred to soluble CD14 [70], whereby lipid A moiety of LPS permits LPS to bind with TLR4 and then activate MyD88 pathway as explained previously [3,47]. In addition, the arrangement of LPS genes into clusters and operons does not always parallel the biosynthetic pathway. The genes in rfa cluster such as *rfaK* and *rfaL* are involved in the synthesis and modification of LPS core. These two genes play a vital role in the attachment of 0 antigens to the core. This result was following the higher protein abundance of LPS (**Figure 2D**) and mRNA expression of LPS biosynthesizing genes *rfaK* and *rfaL* (**Figure 2E**) in cecal tissues of IHF meat geese. Next, to identify whether lipid A moiety of LPS permits LPS to bind with TLR4 and then activate MyD88 pathway, first, we determined the genes related to lipid A biosynthesis. We found that the mRNA expression of *lpxA*, *lpxB*, *lpxC*, and *lpxD* was lesser in the AGF meat geese compared with IHF meat geese (**Figure 2F**). Next, we found decreased mRNA expression of *LBP* (**Figure 2G**) and *soluble cluster of differentiation 14 (sCD14)* (**Figure 2H**). This suggests that the higher mRNA expression of genes related to lipid A may able the LPS to attach with *TLR4* (**Figure 2I**) and then activate MyD88 dependent pathway in cecal tissues of IHF meat geese compared with AGF meat geese at 45d, 60d, and 90d (**Figure 2J**). The activation of TLR4/MyD88 pathway may contribute to ROS production [71]. As expectedly, the higher *TLR4/MyD88* gene expression was observed to be increased in IHF meat geese concerning higher ROS production (**Figure 2K**).

### 3.3. Inhibitory Effects of Artificial Pasture Grazing System on In-house Feeding System Deteriorated Nutrient Absorption

Accumulating studies evince that oxidative stress owing to excessive ROS production either by heat stress or feed may modulate nutrient absorption in broilers [1]. This may happen due to the inability of intestinal ALP to detoxify LPS which decreases the villus height to crypt depth ratio [72,73]. In our study, we found that intestinal ALP activity (Majorbio i-Sanger cloud platform (http://rdp.cme.msu.edu/)) was decreased in IHF meat geese along with increased ROS abundances (**Figure 2K**) compared with AGF meat geese. Based on this scenario, next, to identify whether the IHF-induced intestinal dysbiosis may impact intestinal morphology in meat geese, we performed hematoxylin and eosin (H&E) staining of cecal tissues. The effect of AGF system on the morphology of cecal tissues is presented in **Supplemental Table 3**. Irrespective of a commercial diet, the villus height, villus width, surface area, and distance between villi in cecum of AGF meat geese group were improved as opposed to crypt depth compared to those of IHF meat geese group at different time points 45d, 60d, and 90d. While the villus height to crypt depth ratio (V:C) values of the cecum were not different. Further, the morphology of the cecal tissues from different feeding systems was measured and compared to one another as shown in **Supplemental Figure 6**. These results illustrated the nutrient absorption in cecum under different feeding systems.

### 3.4. Beneficial Effects of Artificial Pasture Grazing System on In-house Feeding System-dependent Apoptosis-induced Gut Permeability in Meat Geese

Recent studies demonstrated that ROS plays a pivotal role in apoptosis [74]. Apoptosis is a key process of programmed cell death which is accompanied by B-cell lymphoma-2 (Bcl-

2) family members that contribute to regulating the permeability of the outer mitochondrial membrane [75]. Alterations in mitochondrial membrane permeability could initiate the stimulation of cytochrome C into the cytoplasm, which activates caspases that, in turn, trigger apoptosis. Before starting apoptosis-related experiments, we confirmed that cytochrome C activity was increased with a commercial diet in the cecal tissues of IHF meat geese (Majorbio i-Sanger cloud platform (http://rdp.cme.msu.edu/)), then further we confirmed it from the mRNA expression of *cytochrome C* from cecal tissues (**Figure 3A**). To discover that apoptosis production is affected by feed type in meat geese, we tested mRNA expression levels of *caspase 3* and *8* in cecal tissues from IHF and AGF meat geese at different stages such as 45d, 60d, and 90d. We found a significant increase in *caspase3* (*CASP3*) and *8* (*CASP8*) activity in IHF meat geese (**Figures 3B and C**). Next, we performed H&E staining of cecal tissues. The upper normal limit for the number of apoptotic cells per field (Mean±SD) in the villus at 45d, 60d, and 90d from the cecal tissues of IHF and AGF meat geese has been shown in **Figure 3D and Supplemental Table 4**. From different studies, this accelerated apoptosis has been involved in inducing intestinal mucosa disruption and intestinal permeability [76,77]. To evaluate whether the mucus phenotype can be explained by altered structural organization of the mucosal barrier, we stained cecal tissues with H&E. A well-defined mucus-producing goblet cell genes (mucin2 (*MUC2*) and Mucin 5, subtype AC (*MUC5AC*)), the number of goblet cells per 20µm, and inner muscular tonic/muscularis mucosa layer thickness were observed in AGF meat geese group (**Figures 3E** and **F; Supplemental Figures 7A; 8A** and **B** and **Supplemental Table 5**). In agreement with a mucosal barrier, the mucus-producing goblet cell genes and inner muscular tonic/muscularis mucosa layer appeared less organized upon a commercial diet feeding. Furthermore, intestinal ALP is known to promote gut barrier function, and the disruption of the intestinal barrier is thought to play a critical role in gut permeability development, hence, we measured the gut permeability in IHF and AGF meat geese at 45d, 60d, and 90d. The ELISA kit method showed an IHF-dependent increase in endotoxemia (LPS) (**Figure 2D**), significantly influenced by intestinal ALP deficiency in IHF meat geese (previously explained in **Figures 2A-C**). Furthermore, expression levels of intestinal tight junction proteins were measured in serum samples of IHF and AGF meat geese. Again diet and loss of serum ALP were associated with a significant reduction in protein expression levels of zona occludin-1 (ZO-1), Occludin, and Claudin (**Figures 3G-I**).

### 3.5. Inhibitory Effects of Artificial Pasture Grazing System on In-house Feeding System-induced NF-kB Pathway and its Systemic Inflammation

In our study first, we explained that the activation of MyD88 dependent pathway could lead to the stimulation of ROS in IHF meat geese. Along with this dependent pathway, the cellular protein LC8 (8-kDa dynein light chain) plays a role in the redox regulation of NF-kB pathway. Actually, LC8 binds to *IkB-α* in a redox-dependent manner, thereby preventing its phosphorylation by IKK [78]. Here, IHF system-induced intestinal ALP disruption and the resulted ROS production oxidized LC8 which leads to the dissociation from *IkB-α* and then causes *NF-kB* activation. This result was under reduced mRNA expression levels of *LC8* and *IkB-α* and increased mRNA levels of *NF-kB* in IHF meat geese (**Figures 4A-C**). Further, we investigated the genes related to regulating NF-kB pathway. The mRNA expression of *NF-kB*-regulated genes *IL-8, CCL2, PLAU*, and *BIRC3* in IHF meat geese (**Figure 4D**) showed that *IL-8*, *PLAU*, and *BIRC3* were by far the most abundant contributor genes involved in regulating the NF-kB pathway in the cecal tissues of IHF meat geese. Similarly, a significant increase in mRNA expression of pro-inflammatory mediators *iNOS* and *COX2* and cytokines *IL-1B, IL-6*, and *TNF-α* in cecal tissues of IHF meat geese were observed instead of *IL-1B* as a function of diet composition. Here we observed that intestinal ALP deficiency was associated with significantly increased mRNA expression levels of these five cytokines (**Figures 4E-I**).

### 3.6. Effects of Artificial Pasture Grazing System on ROS-directed KEAP1 Inhibition and Activation of Nrf2 Pathway in Meat Geese

When NF-kB pathway is established under the insults of ROS, then a natural immune defense mechanism is activated underpinning the Keap1-Nrf2 pathway. *Keap1* regulates the activity of *Nrf2* and acts as a sensor for oxidative stress. Upon oxidative stress, *Keap1* loses its ability to ubiquitinate *Nrf2*, allowing *Nrf2* to move in the nucleus and activate its target genes [79]. In our study, we found that the mRNA expression levels of *Keap1* declined and mRNA expression levels of *Nrf2* were increased by a limited ROS production in AGF meat geese (**Figure 5A**). This demonstrates that the production of ROS activates Nrf2 pathway. Because Nrf2 pathway is known to promote cellular redox homeostasis, and as the impairment of *Nrf2* activity is considered to play a crucial role in cellular defense system, we measured *Nrf2* and *Nrf2*-regulated genes and the antioxidant enzymes regulated by *Nrf2* in IHF and AGF meat geese at three-time points i.e. 45d, 60d, and 90d. With AGF system, the mRNA expression levels of *Nrf2* (**Figure 5B**) and *Nrf2*-regulated genes *NQO1, Gclc, Gclm*, and *GSTA4* were increased in AGF meat geese (**Figure 5C**). Next, we examined that the protein levels of antioxidants HO-1, GSR, T-SOD, GSH-PX, T-AOC, and CAT were also increased with improved *Nrf2* activity in serum samples of AGF meat geese at 45d, 60d, and 90d (**Figures 5D-I**). Further, whether these antioxidants are involved in attenuating the mediators that caused ROS insults in IHF and AGF meat geese, we measured oxidative mediator MDA from the serum samples. We examined a severe increase in protein levels of MDA in IHF meat geese compared with AGF meat geese at three time points 45d, 60d, and 90d (**Figure 5J**).

### 3.7. Artificial Pasture Grazing System Attenuates the In-house Feeding System-induced Endotoxemia, Gut Permeability, and Chronic Systemic Inflammation of Meat Geese

To further explore the effects of long-term establishment of AGF system on intestinal alterations at three-time points 45d, 60d, and 90d, we hypothesize whether intestinal ALP may involve in activating Nrf2 pathway, and we performed correlation analysis among host markers (**Figure 9B**). The results obtained from this relationship showed that intestinal ALP was positively correlated with *Nrf2* and *Nrf2*-regulated genes and as well as its antioxidant enzymes. Further Nrf2 pathway including its antioxidation immune system was positively correlated with *IL-4*, *IL-10*, and tight junction proteins including 2 genes encoding tight junction proteins *Discs large 1 (dlg1)* and *E-cadherin*. IL4 and IL-10 are known to be anti-inflammatory cytokines [80]. Indeed, we also measured the protein levels of cecal ALP, endotoxin (LPS), ROS, ZO-1, Occludin, and Claudin by ELISA kit method (**Figures 6A-F**) and mRNA expression of anti-inflammatory cytokines (*IL-4* and *IL-10*) (**Figures 6G** and **H**), 2 genes encoding tight junction proteins *dlg1* and *E-cadherin* (**Figures 6I** and **J**), and pro-inflammatory cytokines (*iNOS*, *COX2*, *IL-1B*, *IL-6*, and *TNF-α*) (**Figures 6K-O**) that has received AGF and IHF environment from 45d to 90d. The results obtained from correlation analysis suggest that the activation of Nrf2 pathway by intestinal ALP enzyme was primarily involved in attenuating endotoxemia, gut permeability, and pro-inflammatory cytokines in AGF meat geese.

### 3.8. Long-term Artificial Pasture Grazing System Attenuates the Manifestation of KEAP1-induced Aging Phenotypes in Meat Geese

The investigation of impact of Nrf2 pathway activation on cecal tissues aging from the expression levels of aging marker genes showed elevated level of *p19ARF, p16INK4α*, and *p21* in IHF meat geese compared with cecal tissues of AGF meat geese at 45d, 60d, and 90d of age (**Figures 7A-C**). The results illustrated that Nrf2 pathway activation induced by ROS-directed

*Keap1* inhibition effectively suppressed the manifestation of aging phenotypes in cecal tissues of AGF meat geese.

### 3.9. Long-term Artificial Pasture Grazing System Improved Metabolic Profile in Meat Geese

The AGF system was effective in preventing metabolic syndrome in AGF meat geese with a significantly improved body weight (**Figure 8A**) and lipid profile (**Figures 8B-E**), as well as lowering blood glucose and urea nitrogen levels (**Figures 8F and G**).

### 3.10. Long-term Supplementation of Artificial Pasture Grazing System Impedes Compositional Changes in Gut Microbiota by Stimulating Intestinal ALP Enzyme

The microbiota in meat geese' cecal chyme samples were analyzed at three time points (45d, 60d, and 90d) by deep sequencing of the bacteria 16S rRNA gene V3 – V4 region. As shown in **Figure 9A**, the Spearman correlation between microbiota and metabolic indices in the gut tract of meat geese at 45d showed significant positive association between *Bacteroides* with serum GSH-PX, T-AOC, ZO-1, Occludin, and cecal *IL-4, IL-10, LC8*, and *IKB-α* and negative association with serum LPS, ROS, T-AOC, and LDL-C, and cecal *NF-kB, TLR4, MyD88, COX2, IL-6, TNF-a, cytochrome C, CASP3, P16INK4α, P21*, body weight, and blood glucose. *Alistipes* were strongly positively correlated with the cecal pH, serum ALP, HO-1, GSH-PX, T-AOC, and Occludin, and cecal *IL-4, IL-10*, and *LC8* and negatively correlated with serum LPS, MDA, TCHO, TG, LDL-C, and ROS, and cecal *MyD88, NF-kB, TNF-a, cytochrome C, CASP3, p19ARF, p16INK4a, p21*, and blood glucose. *Lactobacillus* was positively correlated with serum ALP, ZO-1, Occludin, and HDL-C, and cecal *Nrf2, IL-4*, and *E- cadherin* and negatively correlated with serum LPS and TG, and cecal *LBP, sCD14, NF-kB, iNOS, IL-1B, cytochrome C, CASP3, Keap1, p19ARF, p16INK4a*, and *p21*. *Norank_f__norank_o__Gastranaerophilales* was positively correlated with serum T-AOC and Occludin and cecal pH and *IL-4* and negatively correlated with serum LPS and cecal *iNOS, CASP3, Keap1, p19ARF*, and *p21*. *Alistipes* and *Lactobacillus* were strongly positively correlated with intestinal ALP and Nrf2 pathway and suppress all those bacteria (*Subdoligranulum, norank_f__norank_o__Clostridia_UCG_014*, and *Erysipelatoclostridium*) that were the causative factors for pathogenesis in AGF meat geese.

As diet has a major influence on gut microbiota composition, richness, and diversity [81]. It has been known that the origin, type, and quality of diet modulate the gut microbiota in a time-dependent manner. Based on the previous studies, we hypothesized whether AGF system as a high dietary fiber source modulates the gut microbiota at different time points or not. We further analyzed the correlation between microbiota and metabolic indices in the gut tract of meat geese at 60d. The results of which showed that the *norank_f__norank_o__RF39* was significantly positively correlated with the cecal pH, serum HO-1, Zo-1, Occludin, claudin, and GSH-PX and cecal *Nrf2, IL-4, IL-10, dlg1*, and *E-cadherin* and negatively correlated with serum LPS and TG and cecal *MyD88, COX2, TNF-a, Keap1, p19ARF*, and *p16INK4a* in AGF meat geese. *Romboutsia* was positively correlated with cecal *Nrf2, IL-4, IL-10, dlg1, E-cadherin*, and *LC8* and negatively correlated with cecal *MyD88, iNOS, TNF-a*, and *p16INK4a*. *Norank_f__norank_o__RF39* and *Romboutsia* were positively correlated with Nrf2 pathway that was strongly involved in attenuating the harmful impacts of *Peptococcus* and *Ruminococcus_torques_group* in AGF meat geese.

To further illustrate the impacts of long-term establishment of AGF system on gut microbial modulation with different time points, we further started analyzing the correlation between microbiota and metabolic indices in the gut tract of meat geese at 90d. The results based on microbial alterations showed that *Faecalibacterium* was positively correlated with cecal pH, GSR, GSH-PX, and T-AOC and cecal *IL-10* and *LC8* and negatively correlated with serum ROS, MDA, and TG, and cecal *LBP*, *TLR4*, *MyD88*, *TNF-a*, *p19ARF*, and *p16INK4a*, and body weight. *Norank_f__Eubacterium_coprostanoligenes_group* was positively correlated with serum HO-1, T-SOD, CAT, ZO-1, Occludin, and HDL-C and cecal *IL-4*, *E-cadherin*, and *IKB-a* and negatively correlated with serum LPS, ROS, and LDL-C and cecal *sCD14*, *NF-kB*, *IL-1B*, *CASP3*, *CASP8*, *Keap1*, and *p21* and blood glucose. At 90 day we observed that *Faecalibacterium* was less positively correlated with ALP that detained the pathogenic effects of *Christensenellaceae_R-7_group* and *Rikenellaceae_RC9_gut_group* in AGF meat geese.

Next, a series of correlation analyses among endotoxemia, gut permeability, pro- and anti-inflammatory cytokines, aging phenotypes, and metabolic syndrome was shown by a Pearson's correlation heat map (**Figure 9B**) in meat geese at 45d, 60d, and 90d. Among them, *LBP* (P = 9.8E-08, R = 0.97351), *sCD14* (P = 1.4E-05, R = 0.92743), *TLR4* (P = 1E-08, R = 0.98324), *MyD88* (P = 3.2E-08, R = 0.97892), ROS (P = 4.6E-07, R = 0.96388), *NF-kB* (P = 3.4E-08, R = 0.97865), *cytochrome C* (P = 2.5E-07, R = 0.968042101), and *Keap1* (P = 2.8E-09, R = 0.98707) were significantly positively correlated with LPS at 45d in IH feeding meat geese. ALP (P = 0.00027, R = -0.8663), *Nrf2* (P = 2.7E-08, R = -0.9796), *IL-4* (P = 3E-07, R = -0.9669), *IL-10* (P = 2.8E-05, R = -0.9164), *dlg1* (P = 0.02022, R = -0.6572), *E-cadherin* (P = 4.4E-08, R = -0.9775), ZO-1 (P = 1.4E-06, R = -0.9546), Occludin (P = 2.6E-05, R = -0.9177), Claudin (P = 4.5E-06, R = -0.9426), *LC8* (P = 9.2E-06, R = -0.9334), and *IKB-a* (P = 1.5E-08, R = -0.9819) were negatively associated with LPS at 45d in GL feeding meat geese.

At 60d *LBP* (P = 1E-08, R = 0.98321), *sCD14* (P = 0.00011, R = 0.88818), *TLR4* (P = 0.00248, R = 0.78524), *MyD88* (P = 1.9E-11, R = 0.99523), ROS (P = 1.6E-09, R = 0.9885), *NF-kB* (P = 4.8E-07, R = 0.96351), *cytochrome C* (P = 0.00016, R = 0.88005), and *Keap1* (P = 1.2E-08, R = 0.98258) were significantly positively correlated with LPS in IH feeding meat geese. ALP (P = 9.2E-08, R = -0.9739), *Nrf2* (P = 1.3E-12, R = -0.9972), *IL-4* (P = 8E-12, R = -0.996), *IL-10* (P = 5E-13, R = -0.9977), *dlg1* (P = 5E-12, R = -0.9964), *E-cadherin* (P = 1.8E-13, R = -0.9981), ZO-1 (P = 5.1E-10, R = -0.9998), Occludin (P = 2.1E-08, R = -0.9806), Claudin (P = 5.5E-11, R = -0.9941), *LC8* (P = 1.6E-10, R = -0.9927), and *IKB-a* (P = 9.4E-07, R = -0.9582) were negatively associated with LPS at 60d in GL feeding meat geese.

At 90d *LBP* (P = 3.2E-10, R = 0.99164), *sCD14* (P = 0.01474, R = 0.68113), *TLR4* (P = 1E-10, R = 0.99336), *MyD88* (P = 4.8E-12, R = 0.99638), ROS (P = 4.6E-10, R = 0.99098), *NF-kB* (P = 2.1E-08, R = 0.98056), *cytochrome C* (P = 5.8E-09, R = 0.98501), and *Keap1* (P = 2.4E-09, R = 0.98744) were significantly positively correlated with LPS in IH feeding meat geese. ALP (P = 1E-08, R = -0.9832), *Nrf2* (P = 2.2E-11, R = -0.9951), *IL-4* (P = 7.2E-12, R = -0.9961), *IL-10* (P = 9.6E-11, R = -0.9934), *dlg1* (P = 6.7E-10, R = -0.9903), *E-cadherin* (P = 1.3E-11, R = -0.9956), ZO-1 (P = 4.5E-12, R = -0.9964), Occludin (P = 5.5E-06, R = -0.94), Claudin (P = 1.3E-05, R = -0.9285), *LC8* (P = 9.1E-11, R = -0.9935), and *IKB-a* (P = 2.2E-08, R = -0.9803) were negatively associated with LPS at 60d in GL feeding meat geese.

## 4. Discussion

Intracellular ROS production by diet-induced gut microbiota facilitated LPS generation [82,83] may lead to chronic low-grade inflammation and modern chronic inflammatory diseases [84,85]. Discovering a safe and novel means of limiting its development is urgently required for the prevention and treatment of these diseases. Diet is a primordial need for life and today, the modern poultry industry is based on grains with lower content of dietary fiber [86]. Moreover, the worldwide trends of excessive low dietary fiber intake have been implicated in today's chronic inflammatory diseases including diabetes mellitus, autoimmune, cancer, cardiovascular, and chronic kidney disease [87,88]. However, the connections between the shifts in dietary fiber contents and attenuating the incidence of chronic inflammatory diseases by activating the intestinal ALP-dependent-redox signaling mechanism remain to be elucidated in meat geese. Therefore the present study demonstrates for the first time, that a long-term pasture grazing can improve commercial diet-induced gut microbial dysbiosis, gut barrier dysfunction and integrity, inflammatory diseases, aging phenotypes, and metabolic syndrome.

Though recent dietary supplementation studies have addressed some impacts of dietary fiber on gut microbiota [88,89]. However, despite these captivating findings, the mechanisms underlying these diverse associations and their outcomings have not been fully explored. In our study, we established artificial pasture grazing system for meat geese that supports concurrent processes to impede in-house feeding system-induced metabolic endotoxemia and systemic inflammation. Consequently, the innovative feature of the AGF system as a high dietary fiber source is to build a pathway-based mechanism by which it increased the abundance of intestinal ALP-producing bacteria, and prevented IHF system-induced LPS-producing bacteria. These changes improve the intestinal nutrient absorption, mucus layer, and mucus-producing goblet cell genes, resulting in reduced metabolic endotoxemia (LPS), LPS-induced ROS production, and gut permeability. The subsequent reduction of pro-inflammatory cytokines leads to the prevention of chronic inflammatory diseases and aging phenotypes. Spearman and Pearson's correlation analysis, including the above-mentioned findings, strongly supports the proposed mechanisms.

Metabolic endotoxemia can be determined by the abundance of bacteria affecting LPS production [22]. In this study, we found that AGF intervention reduced the enrichment of genes involved in LPS biosynthesis based on the predicted function by 16S rRNA sequencing and PICRUSt analysis. Interestingly, our current findings have shown similar results with AGF intervention similar to dietary capsaicin in human subjects [90]. This would indicate the possibility that lower abundance of Gram-negative microbiota must be responsible for the low abundance of COG orthology belonging to LPS biosynthesis functions in the AGF treatment meat geese. This could mainly be due to the prevention of members of the Gram-positive phyla *Firmicutes* (genera *Lactobacillus, Ruminococcus_torques_group*, *Subdoligranulum*, and *Christensenellaceae_R-7_group*) and *Actinobacteriota* (genera *norank_f_norank_o_Gastranaerophilales*) and gram-negative phyla *Bacteroidota* (genera *Prevotellaceae_UCG_001*, *Bacteroides*, *Alistipes*, and *Rikenellaceae_RC9_gut_group*) with AGF intervention because these were the key bacterial phyla and their respective genera that largely contribute to the IHF meat geese. These lipopolysaccharides are bacterium-associated molecular patterns, which act via TLR4/MyD88 pathway by promoting the inflammatory response [65].

The production of ROS by LPS-induced TLR4/MyD88 pathway activation [71] may depend on a diet rich in high fat, high calorie, high protein, and high carbohydrate [66,91]. The correct cellular response to ROS production is critical to preventing oxidative damage and maintaining cell survival. However, when too much cellular damage has occurred, it is to the advantage of a multicellular organism to remove the cell for the benefit of the surrounding cells. ROS can therefore trigger apoptotic cell death based on the severity of the oxidative stress [92] and may contribute to inducing NF-$\kappa$B pathway [93]. ROS-induced severe apoptotic cell death may accelerate intestinal mucosa disruption and result in causing intestinal permeability [76,77]. Based on the current studies, we hypothesized how intestinal ALP would incinerate in LPS-induced TLR4/MyD88 facilitates ROS production pathway.

Intestinal ALP is a major enzyme of interest for its gut microbiota-modifying properties [37,48]. Endogenous ALP production has been shown to inhibit the overgrowth of *E. coli* by dephosphorylating LPS [22,49,94]. It is well known that intestinal ALP capacity to dephosphorylate LPS was shown to be present in the colon and feces of mice [46] and reduces LPS-induced gut permeability and inflammation in Caco2 and T84 cell lines [28,95]. LPS binds specifically to TLR4 and stimulates inflammation by activating two distinct pathways, namely LPS-dependent release of TNF-α and NF-κB (through MyD88-dependent and –independent pathways) [96]. The data from our experiment support the notion that the AGF system as high dietary fiber source enhanced the abundance of intestinal ALP producing *Alistipes*, and *Lactobacillus*, (45d), *Norank_f__norank_o__RF39* and *Romboutsia* (60d), and *Faecalibacterium* (90d) genera. This microbiota was further seen to involve in suppressing *E. coli* and inactivating the capacity of lipid A biosynthesizing genes (*lpxA*, *lpxB*, *lpxC*, and *lpxD*) to bind LPS with TLR4 and then inhibit the activation ability of MyD88 dependent pathway. Some reports, utilizing dietary fiber as a nutrient source in animals, support our results [42,97].

It is well-known that endogenous ALP production enhances the expression of proteins (Zo-1, Occludin, and Claudin) involved in tight junctions, thereby preventing the translocation of endotoxins (LPS) by intestinal gram-negative bacteria (*E. coli*) across the gut barrier [22,38]. Our results were by the reports of Kaliannan et al. [22]; Schroeder et al. [98]; Kühn et al. [38]; and Mei et al. [99], in which microbially-induced endogenous intestinal ALP production was observed to decrease the ROS production and apoptosis-related genes *CASP3* and *CASP8*, improve the mucus-producing goblet cell genes *MUC2* and *MUC5AC* as well as inner muscular tonic/muscularis mucosal layer thickness. Moreover, tight junction proteins Zo-1, Occludin, and Claudin and 2 genes encoding tight junction protein *dlg1* and *E-cadherin* were observed to be increased in AGF meat geese which gave the fact that these proteins were strongly involved in inducing nutrient absorption and overall intestinal health (**Figure 10**).

ROS production regulates *NF-κB* activity in a bidirectional fashion, namely, ROS may trigger activation or repression of *NF-κB* activity [100]. In many studies, *NF-κB* inhibition is LC8 dependent [78]. But in our study, we have shown that the activation of *NF-κB* by LPS-induced TLR4/MyD88-accelerated ROS production is collectively *LC8* and *IkB-a* dependent. Some reports, utilizing LPS as a ROS inducer, support our results [101,102]. The results of our study with a little modification from those of Cario [103] and Fukata et al. [96], evinced that the triggering of LPS facilitated MyD88 pathway and the subsequent ROS production may altogether activate the *NF-κB* signaling cascades, which played an important role in the development of inflammation by synthesizing and stimulating pro-inflammatory cytokines (*iNOS, COX2, IL-1β, IL-6,* and *TNF-α*) [50,104]. Dysregulated inflammatory cytokines production plays a pivotal role in developing low-grade inflammation [105]. The above-mentioned so-called studies were unable to describe whether the activation of NF-κB pathway and the resulted pro-inflammatory cytokines were owing to connections between the dietary components (dietary fiber) and gut microbiota. Our results were in accordance with the report of Kyung-Ah Kim et al. [106]; Eva d'Hennezel et al. [6]; and Sun et al. [85] in which the microbially-induced LPS production increased the mRNA expression of *NF-κB, iNOS, COX2, IL-1β, IL-6,* and *TNF-α* in the liver of mice, pigs, and humans as that of our IHF treatment meat geese. Conversely, in our study, we discovered from the spearman correlation analysis between microbiota and host markers that AGF intervention prevented the IHF-induced upregulation of gut microbiota interacting with LPS, ROS, and pro-inflammatory cytokines by activating gut microbiota those directly interacting with intestinal ALP and Nrf2 pathway. Further, following the mechanism of Bates et al. [107] and Estaki et al. [49], we developed combined Pearson's correlation analysis among host markers to evaluate whether intestinal ALP is involved in activating Nrf2 pathway. We observed that the intestinal ALP was significantly positively correlated with Nrf2 in all stages of sample collection suggesting that intestinal ALP may contribute to activating Nrf2 pathway in AGF treatment meat geese.

In response to oxidative challenge, a stress response is activated to control ROS overproduction and provide optimal conditions for effective ROS signaling to support redox homeostasis. The Nrf2/Keap1 and nuclear factor kappa-light-chain-enhancer of activated B cells/inhibitory κB protein (NF-κB/IkB) systems were considered to be two major "master regulators" of the stress response. One of the most important ways in which *NF-κB* activity influences ROS levels is via increased expression of antioxidant proteins has been explained elsewhere [108]. The way by which Keap1-Nrf2 responds to ROS has not been elucidated clearly in meat geese. Upon oxidative stress, *Keap1* acts as a sensor and regulates the activity of *Nrf2* thereby, *Keap1* loses its ability to ubiquitinate *Nrf2*, allowing *Nrf2* to move in the nucleus and activate its target genes [79,109,110]. The results of our study following this mechanism by which dietary fiber in AGF meat geese was able to activate intestinal ALP that was significantly positively correlated with *Nrf2* and *Nrf2*-regulated genes including *NQ01, Gclc, Gclm,* and *GSTA4,* and the antioxidant defense network-related enzymes such as HO-1, GSR, T-SOD, GSH-PX, T-AOC, and CAT and significantly negatively correlated with oxidative related enzyme MDA.

Several studies revealed that *Nrf2* activity is modulated with different dietary interventions such as high-fat diet or dietary energy restriction [111,112]. The results suggest that the progression of age-related phenotypes *p19ARF, p16INK4α,* and *p21* detected in this study are primarily caused by the decline of protective function by Nrf2 pathway in IHF meat geese. This may be because of chronic smoldering inflammation which is considered one of the important factors associated with low fiber diet-related diseases and aging phenotypes [88,113]. We indeed observed that the expression of pro-inflammatory cytokine genes *iNOS, COX2, IL-1β, IL-6,* and *TNF-a* were increased in IHF meat geese. While epidemiological evidence shows that *TNF-a* and IL-6 are predictive of many aging phenotypes [114]. Hence, we found it to be associated with more pronounced aging phenotypes *p19ARF, p16INK4α,* and *p21* in geese lacking pasture intake, whereas long-term AGF system significantly induced the potent anti-inflammatory action of *Nrf2* and it may evolve in reducing the pro-inflammatory cytokines in AGF meat geese with concomitantly, aging phenotypes. Of note, our results suggest that AGF-induced intestinal ALP positively correlates with *Nrf2* and negatively correlates with *Keap1* and pro-inflammatory cytokines. This notion coincides with the fact that *Nrf2*-mediated inhibition of *iNOS, COX2, IL-1β, IL-6,* and *TNF-a* induction contributes to the prevention of delayed aging phenotypes [115] and the development of geese' health.

In a previous study, intestinal ALP regulation prevents and reverses the changes associated with a high-fat diet-induced metabolic syndrome [38]. Furthermore, regulation of intestinal ALP by dietary fiber-rich diets improves the lipid profile during low dietary fiber and low-fat diets [116,117]. In the current study, we found a low dietary fiber-related spontaneous increase in the serum lipid profile and glucose levels in meat geese, significantly more pronounced in geese lacking pasture intake, underscoring the potential beneficial role of AGF-induced intestinal ALP in the prevention of metabolic diseases.

To prove the hypothesis that intestinal ALP might directly contribute to the reduction of endotoxemia, gut permeability, pro-inflammatory cytokines, and metabolic syndrome, we applied a combined correlation analysis among them. We found that ROS production owing to microbially-induced LPS was seen to be increased with IHF system and further involved in inducing intestinal mucosa deterioration, apoptosis, gut permeability, oxidants, NF-κB pathway, pro-inflammatory cytokines, aging phenotypes, and metabolic syndrome. The establishment of AGF system as a high dietary fiber source can reverse this process. Specifically, AGF system increase the abundance of ALP-producing bacteria and that intestinal ALP negatively correlates with ROS. The low production of ROS in AGF meat geese interacts with *Keap1* and diminishes its activity and then alternatively activates the Nrf2 pathway. Activation of intestinal ALP and Nrf2 pathway collectively positively correlates with *LC8*, *IKB-a*, antioxidants (HO-1, GSR, T-SOD, GSH-PX, CAT, and T-AOC), tight junction proteins ZO-1, Occludin, and Claudin, including 2 genes encoding tight junction proteins *dlg1* and *E-cadherin*, and anti-inflammatory cytokines (*IL-4* and *IL-10*). *IL-4* is produced by Th2 cells [118] whereas *IL-10* is involved in Th2 differentiation [119] and both are known to be anti-inflammatory cytokines [80]. Several pieces of evidence from previous studies revealed that *IL-4* and *IL-10* depletion is associated with pronounced ulcerative colitis and Crohn's disease, type 2 diabetes, metabolic syndrome [120-122]. In our study, the activation of intestinal ALP, Nrf2 pathway, antioxidants, *IKB-a*, and anti-inflammatory cytokines potentially evolve in reducing endotoxemia, gut permeability, pro-inflammatory cytokines, aging phenotypes, and metabolic syndrome in AGF meat geese.

In summary, our data suggest that intestinal ALP – as a natural brush border enzyme – plays a critical role in animal health development through maintaining intestinal microbiome homeostasis, reducing LPS-induced ROS production, activating Nrf2 pathway, inducing anti-inflammatory immune responses, and preserving gut barrier function, decreasing low-grade inflammation, and metabolic syndrome. Further studies will focus on elucidating the precise mechanisms of intestinal ALP and Nrf2 pathways' beneficial role in different dietary patterns and aging. Given that AGF system safely induces intestinal ALP and Nrf2 pathways, targeting specific dietary fiber sources that could induce endogenous intestinal ALP production could represent a novel approach to preventing a variety of diet-induced gut microbial-related diseases in animals.

## 5. Conclusions

Microbially-induced ALP production by AGF system appears to preserve intestinal homeostasis by targeting crucial intestinal alterations, including endotoxemia, gut barrier dysfunction, systemic chronic low-grade inflammation, and metabolic syndrome. AGF system as a high dietary fiber source may represent a novel therapy to counteract the chronic inflammatory state leading to low dietary fiber-related diseases in animals.

## 6. Data Availability Statement

The datasets presented in this study can be found in online repositories. The names of the repository/repositories and accession number(s) can be found below:…….

## 7. Author contributions

QA designed research, conducted experiments, acquired data, analyzed data, performed statistical analysis, and wrote the manuscript. JN, FL, MA, BL, SL, DL, ZW, HS, and YC acquired data and conducted experiments. YS, SM, and UF designed research, analyzed data, and critically revised the manuscript for intellectual content. All authors revised and approved the manuscript for publication.

## 8. Funding

This work was supported by grants from the Modern Agro-industry Technology Research System of China (CARS-34) and the Science and Technology Innovation Team of Henan Province High Quality Forage and Animal Health (No.22IRTSTHN022).

## 9. Acknowledgments

We thank Henan Daidai goose Agriculture and Animal husbandry development Co. LTD (Zhumadian, China) for providing Wanfu meat geese stocks used in this study. We thank Shi yinghua and San Ma for assistance with microbiome analysis.

**Figure legends**

**Figure. 1** Artificial pasture grazing system modulates gut microbiota to inhibit LPS synthesis induced by in-house feeding system. (A) Average abundance per sample of genes related to the four main LPS biosynthesis-related functions. Red indicates a positive correlation; green indicates a negative correlation, (B) Relative contributions of the different phyla to the total LPS-encoding capacity of the gut microbiome, and (C) Contribution of individual genera to LPS biosynthesis functions. The average abundances of genes related to any of the four LPS-related GO functions are shown for individual genera within each phylum. LP A biosyn. AC.; lipid A biosynthesis acyltransferase, LP A biosyn.; lipid A biosynthesis, LPS trans. peri. prot. lptA; lipopolysaccharide transport periplasmic protein lptA, and LPS biosyn. proc.; lipopolysaccharide biosynthesis process. Red indicates a positive correlation; blue indicates a negative correlation. In-house feeding system (IHF) and artificial pasture grazing system (AGF). The asterisks symbol indicates significant differences *$P < 0.05$, **$P < 0.01$.

**Figure 2** Inhibitory effects of artificial pasture grazing system on in-house feeding system-induced ROS production via LPS/TLR4/MyD88 pathway in meat geese. (**A**) ALP protein level in serum, (**B**) *ALPi* mRNA level in cecal tissues, (**C**) mRNA levels of intestinal ALP genes (*CG5150* and *CG10827*) in cecal tissues, (**D**) LPS protein level in serum, (**E**) mRNA levels of LPS biosynthesizing genes (*rfaK* and *rfaL*) in cecal tissues, (**F**) mRNA levels of lipid A biosynthesizing genes (*lpxA*, *lpxB*, *lpxC*, and *lpxD*) in cecal tissues, (**G**) *LBP* mRNA level in cecal tissues, (**H**) *sCD14* mRNA level in cecal tissues, (**I**) *TLR4* mRNA level in cecal tissues, (**J**) *MyD88* mRNA level in cecal tissues, and (**K**) ROS protein level in serum, normalized by *β-actin* and measured by qPCR. In-house feeding system (IHF) and artificial pasture grazing system (AGF). Data with different superscript letters are significantly different (P < 0.05) according to the unpaired student T-Test. The asterisks symbol indicates significant differences *$P < 0.05$, **$P < 0.01$.

**Figure 3** Beneficial effects of artificial pasture grazing system on in-house feeding system-dependent apoptosis-induced gut permeability in meat geese. (**A**) *Cytochrome C* mRNA level in cecal tissues, (**B**) *CASP3* mRNA level in cecal tissues, and (**C**) *CASP8* mRNA level in cecal tissues. (**D**) H&E staining of cecal tissues (magnification, 40×). (**E**) *MUC2* mRNA level in cecal tissues, and (**F**) *MUC5AC* mRNA level in cecal tissues, normalized by *β-actin* and measured by qPCR. (**G**) ZO-1 protein level in serum, (**H**) Occludin protein level in serum, and (**I**) Claudin protein level in serum. In-house feeding system (IHF) and artificial pasture grazing system (AGF). Data with different superscript letters are significantly different (P < 0.05) according to the unpaired student T-Test. The asterisks symbol indicates significant differences *$P < 0.05$, **$P < 0.01$.

**Figure 4** Inhibitory effects of artificial pasture grazing system on in-house feeding system-induced NF-kB pathway and its systemic inflammation. (**A**) *LC8* mRNA level in cecal tissues, (**B**) *IKB-a* mRNA level in cecal tissues, (**C**) *NF-kB* mRNA level in cecal tissues, (**D**) NF-kB-regulated genes (*IL-8, CCL2, PLAU,* and *BIRC3*) mRNA levels in cecal tissues, (**E**) *iNOS* mRNA level in cecal tissues, (**F**) *COX2* mRNA level in cecal tissues, (**G**) *IL-1B* mRNA level in cecal tissues, (**H**) *IL-6* mRNA level in cecal tissues, and (**I**) *TNF-a* mRNA level in cecal tissues, normalized by *β-actin* and measured by qPCR. In-house feeding system (IHF) and artificial pasture grazing system (AGF). Data with different superscript letters are significantly different (P < 0.05) according to the unpaired student T-Test. The asterisks symbol indicates significant differences *$P < 0.05$, **$P < 0.01$.

**Figure 5** Effects of artificial pasture grazing system on ROS-directed *KEAP1* inhibition and activation of Nrf2 pathway in meat geese. (**A**) *Keap1* mRNA level in cecal tissues, (**B**) *Nrf2* mRNA level in cecal tissues, and (**C**) Nrf2-regulated genes (*NQO1, Gclc, Gclm,* and *GSTA4*) mRNA levels in cecal tissues, normalized by β-actin and measured by qPCR. (**D**) HO-1 protein level in serum, (**E**) GSR protein level in serum, (**F**) T-SOD protein level in serum, (**G**) GSH-PX protein level in serum, (**H**) T-AOC protein level in serum, (**I**) CAT protein level in serum, and (**J**) MDA protein level in serum. In-house feeding system (IHF) and artificial pasture grazing system (AGF). Data with different superscript letters are significantly different (P < 0.05) according to the unpaired student T-Test. The asterisks symbol indicates significant differences *$P < 0.05$, **$P < 0.01$.

**Figure 6.** Artificial pasture grazing system attenuates the in-house feeding system-induced endotoxemia, gut permeability, and chronic systemic inflammation of meat geese**.** (**A**) ALP protein level in cecal tissues, (**B**) LPS protein level in cecal tissues, (**C**) ROS protein level in cecal tissues, (**D**) ZO-1 protein level in cecal tissues, (**E**) Occludin protein level in cecal tissues, and (**F**) Claudin protein level in cecal tissues. (**G**) *dlg1* mRNA level in cecal tissues, (**H**) *E-cadherin* mRNA level in cecal tissues, (**I**) *IL-4* mRNA level in cecal tissues, (**J**) *IL-10* mRNA level in cecal tissues, (**K**) *iNOS* mRNA level in cecal tissues, (**L**) *COX2* mRNA level in cecal tissues, (**M**) *IL-1B* mRNA level in cecal tissues, (**N**) *IL-6* mRNA level in cecal tissues, and (**O**) *TNF-a* mRNA level in cecal tissues, normalized by *β-actin* and measured by qPCR. In-house feeding system (IH) and artificial pasture grazing system (AGF). Data with different superscript letters are significantly different (P < 0.05) according to the unpaired student T-Test. The asterisks symbol indicates significant differences *$P < 0.05$, **$P < 0.01$.

**Figure 7** Effect of different feeding systems on KEAP1-induced aging phenotypes in meat geese. (**A**) *p19ARF* mRNA level in cecal tissues, (**B**) *p16INK4α* mRNA level in cecal tissues, and (**C**) *p21* mRNA level in cecal tissues, normalized by *β-actin* and measured by qPCR. In-house feeding system (IHF) and Artificial pasture grazing system (AGF). Data with different superscript letters are significantly different (P < 0.05) according to the unpaired student T-Test. The asterisks symbol indicates significant differences *$P < 0.05$, **$P < 0.01$.

**Figure 8** Effect of different feeding systems on metabolic profile of meat geese. (**A**) Body weight (kg), (**B**) T-CHO protein level in serum, (**C**) LDL-C protein level in serum, (**D**) HDL-C protein level in serum, (**E**) TG protein level in serum, (**F**) Blood glucose levels, and (**G**)

BUN protein level in serum. In-house feeding system (IHF) and artificial pasture grazing system (AGF). Data with different superscript letters are significantly different (P < 0.05) according to the unpaired student T-Test. The asterisks symbol indicates significant differences *P < 0.05, **P < 0.01.

**Figure 9.** Association and model predictive analysis of the top 44 host markers with the highest correlation scores. (**A**) Correlation between gut microbiota and host markers by Spearman correlation analysis. Red squares indicate a positive correlation; whereas blue squares indicate a negative correlation. (**B**) The labels on the abscissa and the longitudinal axis represent Pearson's correlation heatmap among host markers. Red squares indicate a positive correlation and blue squares indicate a negative correlation. Deeper colors indicate stronger correlation scores.

**Figure 10. Diagram illustrating a proposed mechanism by which artificial pasture grazing system as a high dietary fiber source up-regulates intestinal ALP-producing bacteria and Nrf2 signaling pathway while downregulating LPS-producing bacteria and ROS in meat geese.** The increase in intestinal ALP-producing bacteria and the activation of Nrf2 signaling pathway maintain antioxidant and anti-inflammatory mechanisms that lower LPS-producing bacteria and LPS-induced ROS generation, intestinal mucosal deterioration, gut permeability, and metabolic endotoxemia. In the first step, the intestinal ALP attack on TLR4 and let the lipid A moiety not to allow LPS to bind with TLR4 and dephosphorylate LPS by breaking TLR4/MyD88-induced ROS production. In the second step, intestinal ALP activates Nrf2 pathway which reduces oxidative stress, so that ROS could not oxidize LC8 protein and deteriorate *IKB-a* to activate NF-κB pathway. In this way, intestinal ALP activates anti-inflammatory cytokines and then attenuates chronic low-grade inflammation, aging phenotypes, and metabolic syndrome. BG: blood glucose; TG: triglyceride; BW: body weight; BUN: blood urea nitrogen.

**Table 1: Nutritional composition of the diet**

| Ingredients, % | Diets | |
| --- | --- | --- |
| | Grower | Finisher |
| Wheat | 57.3 | 59 |
| Rice bran | 5 | 4 |
| Corn germ cake (exp.) | 4 | 3.2 |
| Corn oil | 5 | 7 |
| Dumpling powder | 3 | 2 |
| Corn distiller's grains (DDGS) | 6.5 | 7 |
| Spouting germ meal | 3 | 2 |
| Soybean meal (sol.) | 7 | 6 |
| Peanut meal (sol.) | 1.5 | 1 |
| Albumen powder | 2 | 1.5 |
| Stone powder | 1.1 | 1 |
| Liquid methionine | 0.25 | 0.3 |
| MuLaoDa-2 | 1.25 | 2 |
| 201/202 gunk | 2.5 | 3 |
| Calcium hydrogen phosphate | 0.6 | 1 |
| *Chemical composition (%)* | | |
| Crude protein | 20.12 | 15.54 |
| Ash | 12.89 | 12.86 |
| NDF | 13.25 | 30.55 |
| ADF | 5.5 | 27.02 |
| Ca | 1.15 | 1.07 |
| P | 0.47 | 0.32 |

**Supporting information**

**Supplemental Figure 1.** Overview of feeding and sampling strategies.

**Supplemental Figure 2. Contributions of bacteria at phylum level to LPS biosynthesis functions.** (A-F) Relative abundances (%) of the six most dominant phyla in the cecal chyme of the IHF and AGF meat geese. Data with different superscript letters are significantly different ($P < 0.05$) according to the unpaired student T-Test. *$P < 0.05$, **$P < 0.01$.

**Supplemental Figure 3. Contribution of individual genera within each phylum to LPS biosynthesis functions.** The complete name for each genus is given below. (**A**) Relative abundances (%) of the sixteen most dominant genera (*Subdoligranulum, norank_f__norank_o__Clostridia_UCG-014, unclassified_f__Lachnospiraceae, Lactobacillus, Faecalibacterium, Erysipelatoclostridium, Ruminococcus_torques_group, unclassified_f__Oscillospiraceae, Faecalitalea, Peptococcus, Blautia, Romboutsia, Christensenellaceae_R-7_group, norank_f__Eubacterium_coprostanoligenes_group, Enterococcus, Butyricicoccus,* and *norank_f__norank_o__RF39*) within phylum *Firmicutes* in the cecal contents of the IH and GL feeding meat geese. (**B**) Relative abundances (%) of the five most dominant genera (*Bacteroides, Alistipes, Parabacteroides, Prevotellaceae_UCG-001,* and *Rikenellaceae_RC9_gut_group*) within phylum *Bacteroidota* in the cecal contents of the IH and GL feeding meat geese. (**C-E**) Relative abundances (%) of the most dominant genera within phylum *Actinobacteriota* (*Bifidobacterium*), *Cyanobacteria* (*norank_f__norank_o__Gastranaerophilales*), and *Desulfobacterota* (*Desulfovibrio*) in the cecal chyme of the IHF and AGF meat geese. Data with different superscript letters are significantly different (P < 0.05) according to the unpaired student T-Test. *$P$ < 0.05, **$P$ < 0.01.

**Supplemental Figure 4.** **Effect of different feeding systems on pH values of meat geese gastrointestinal tract.** (**A**) pH of proventriculus, (**B**) pH of gizzard, (**C**) pH of ileum, and (**D**) pH of cecum. In-house feeding system (IHF) and artificial pasture grazing system (AGF). Data with different superscript letters are significantly different (P < 0.05) according to the unpaired student T-Test. The asterisks symbol indicates significant differences *$P$ < 0.05, **$P$ < 0.01.

**Supplemental Figure 5. Effects of different feeding systems on *E. coli* production in cecal tissues of meat geese.** (**A**) Representative culture plate photos showing the difference between IHF and AGF meat geese in the growth of LPS-producing gram-negative *E. coli.* CFU/g stool. (**B**) *E. coli* cell cultures based on spectrophotometer readings at OD600 for 48hrs. Data with different superscript letters are significantly different (P < 0.05) according to the unpaired student T-Test. *$P$ < 0.05, **$P$ < 0.01.

**Supplemental Figure 6.** Effects of different feeding systems on cecal morphology (100μm). VH – villus height; VW – villus width; DBV – distance between two villi; CD – crypt depth. In-house feeding system (IHF) and artificial pasture grazing system (AGF).

**Supplemental Figure 7.** (**A**) Comparison of the goblet cell number (per 20μm) of meat geese with different feeding systems. H&E staining of cecal tissues (magnification, 40×). Goblet cell (GC), In-house feeding system (IHF) and artificial pasture grazing system (AGF). Data with different superscript letters are significantly different (P < 0.05) according to the unpaired student T-Test. The asterisks symbol indicates significant differences *$P$ < 0.05, **$P$ < 0.01.

**Supplemental Figure 8.** (**A**) Light micrograph of the wall of cecum tissues of meat geese, hematoxylin and eosin (40μm): *1* – outer layer of muscular tonic; *2* – inner layer of muscular tonic; *3* – outer layer of lamina muscularis mucosa (LMM); *4* – submucosal nerve node; and *5* – inner layer of lamina muscularis mucosa (LMM). (**B**) Comparison of the cecal membrane thickness of meat geese with different feeding systems (50μm). In-house feeding system (IHF) and artificial pasture grazing system (AGF).

# Figures

## Figure. 1

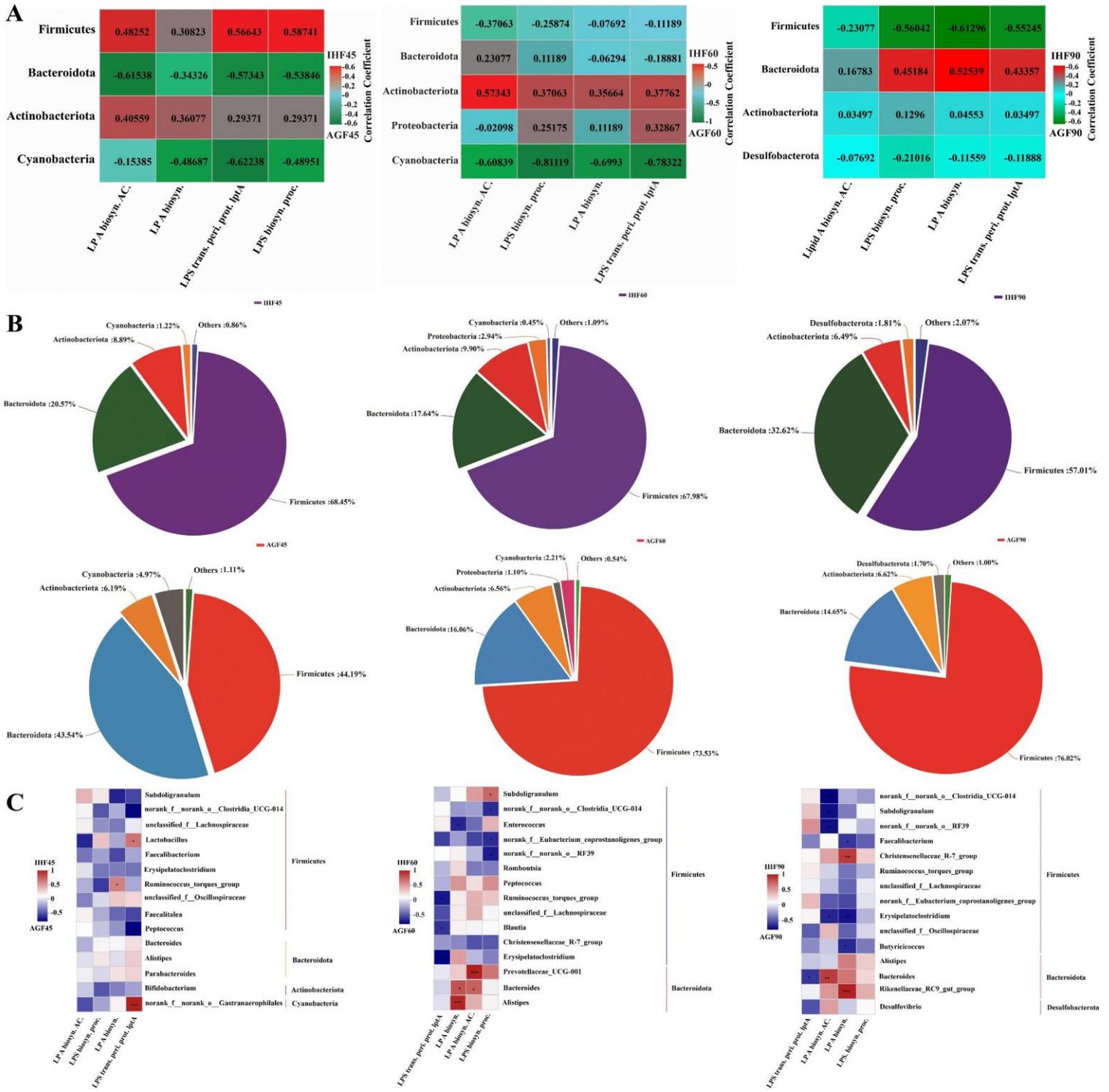

**Figure. 2**

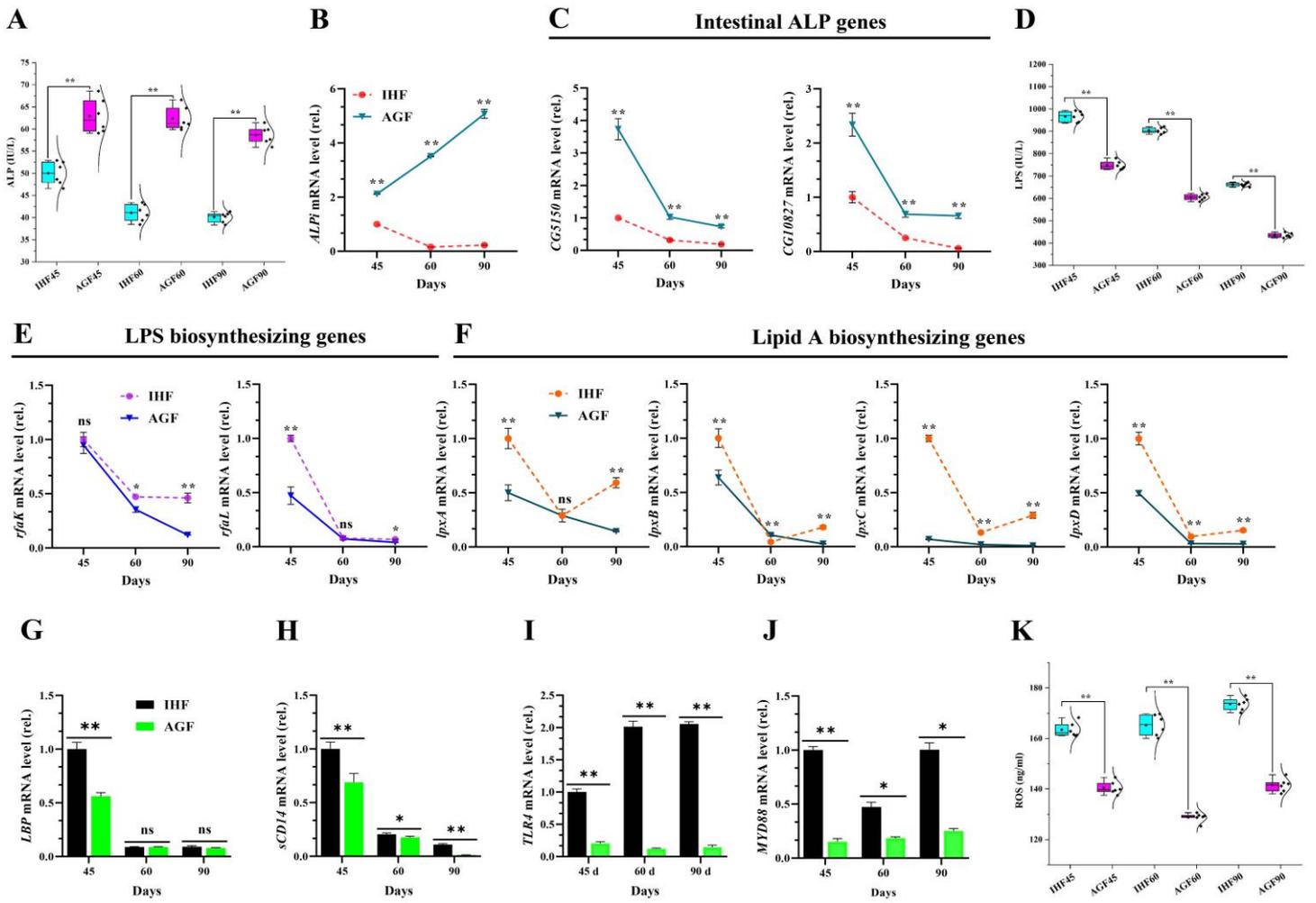

**Figure. 3**

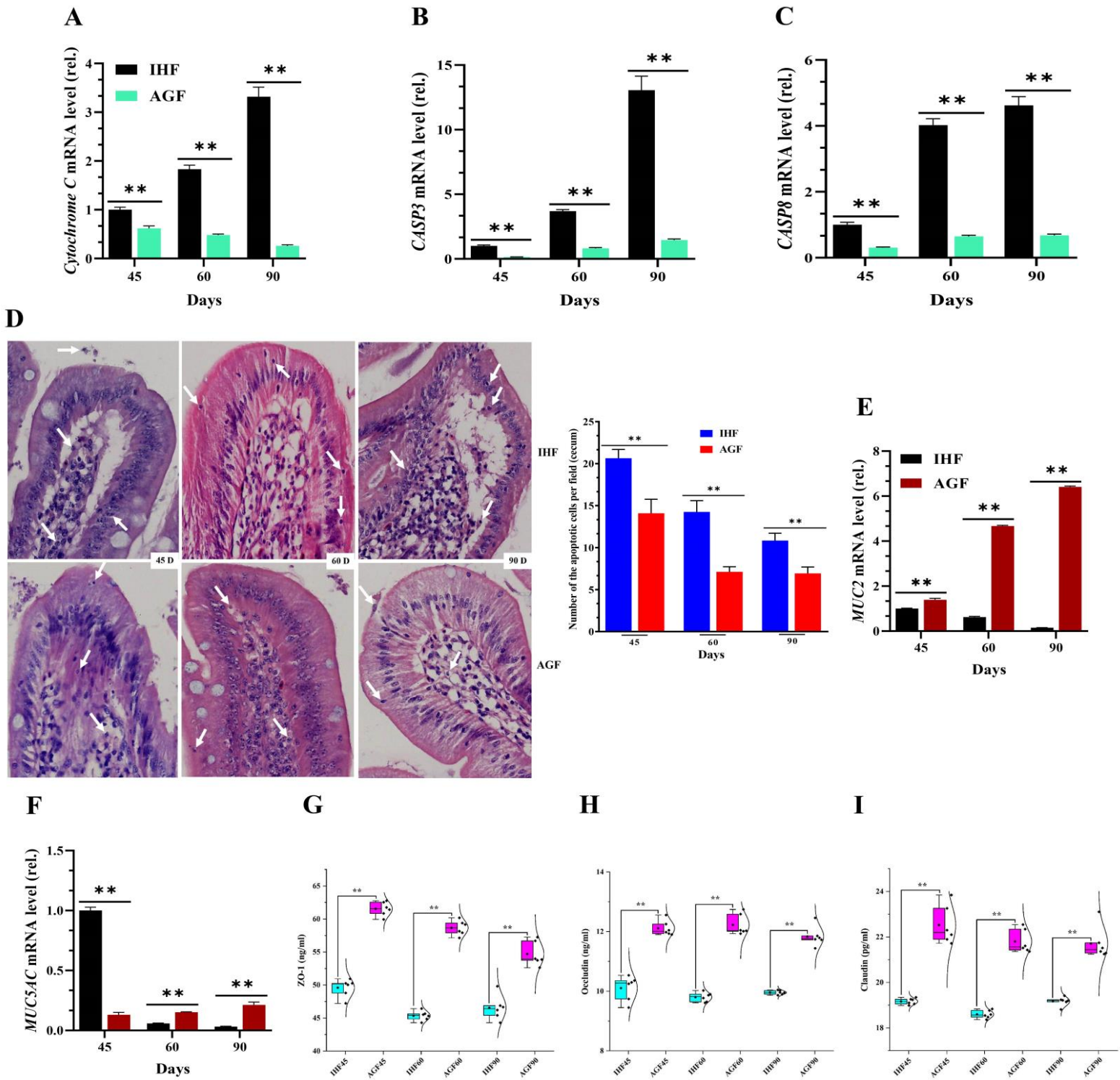

**Figure. 4**

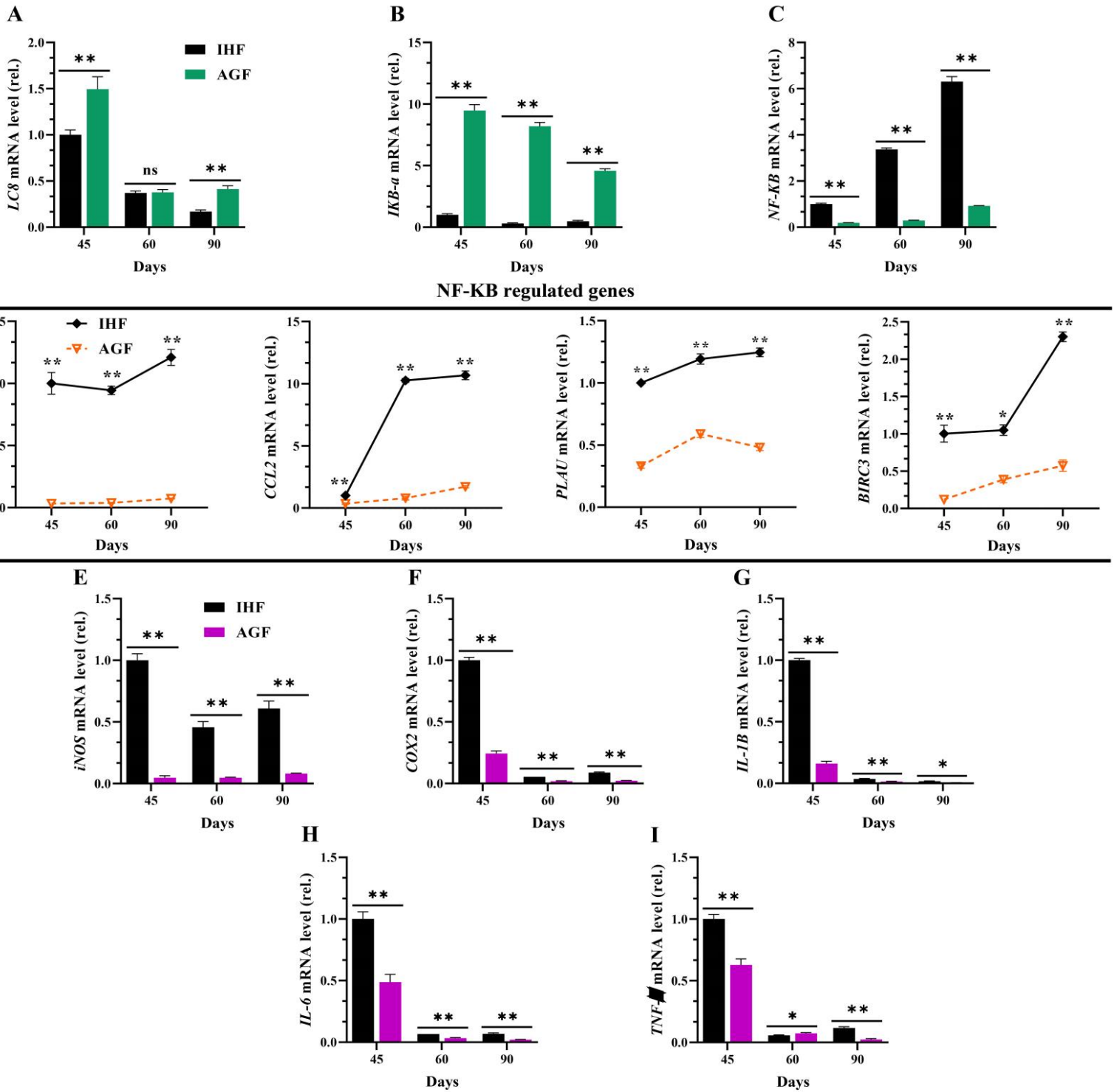

**Figure. 5**

**A**

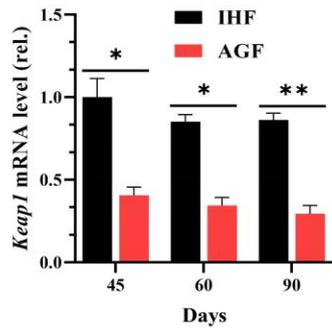

**B**

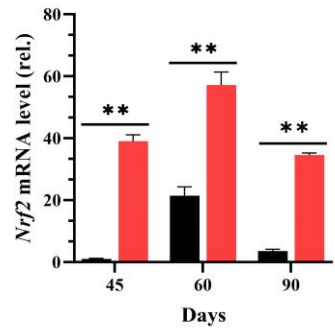

**C** Nrf2 regulated genes

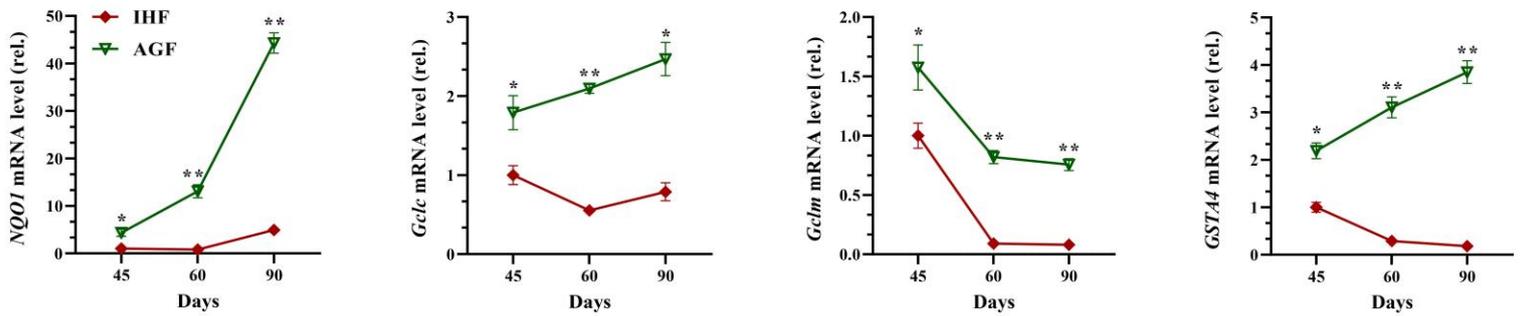

**D**

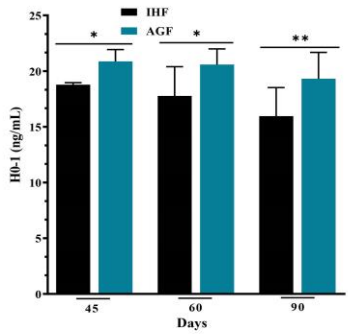

**E**

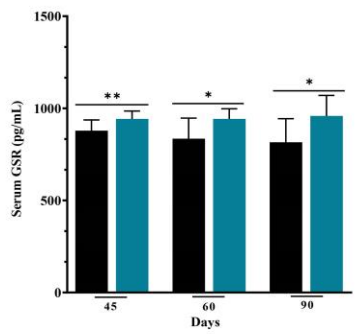

**F**

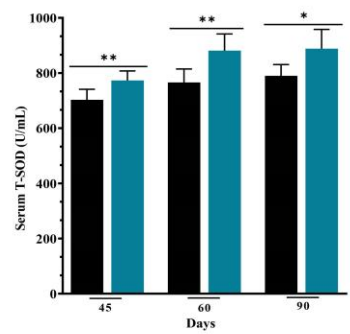

**G**

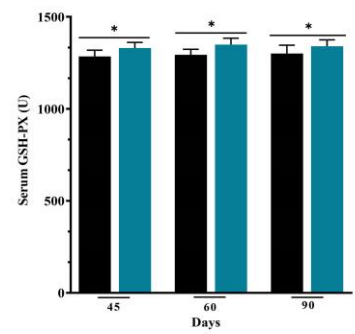

**H**

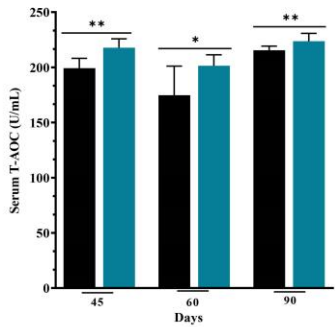

**I**

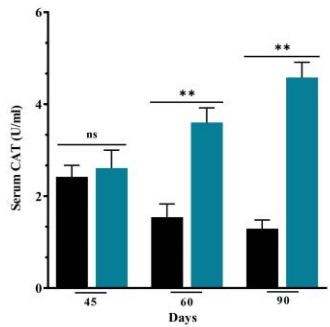

**J**

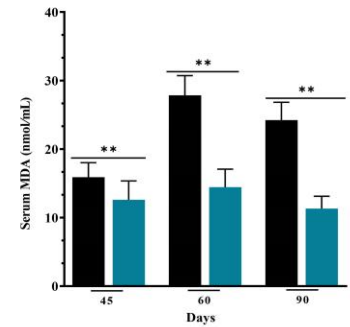

**Figure. 6**

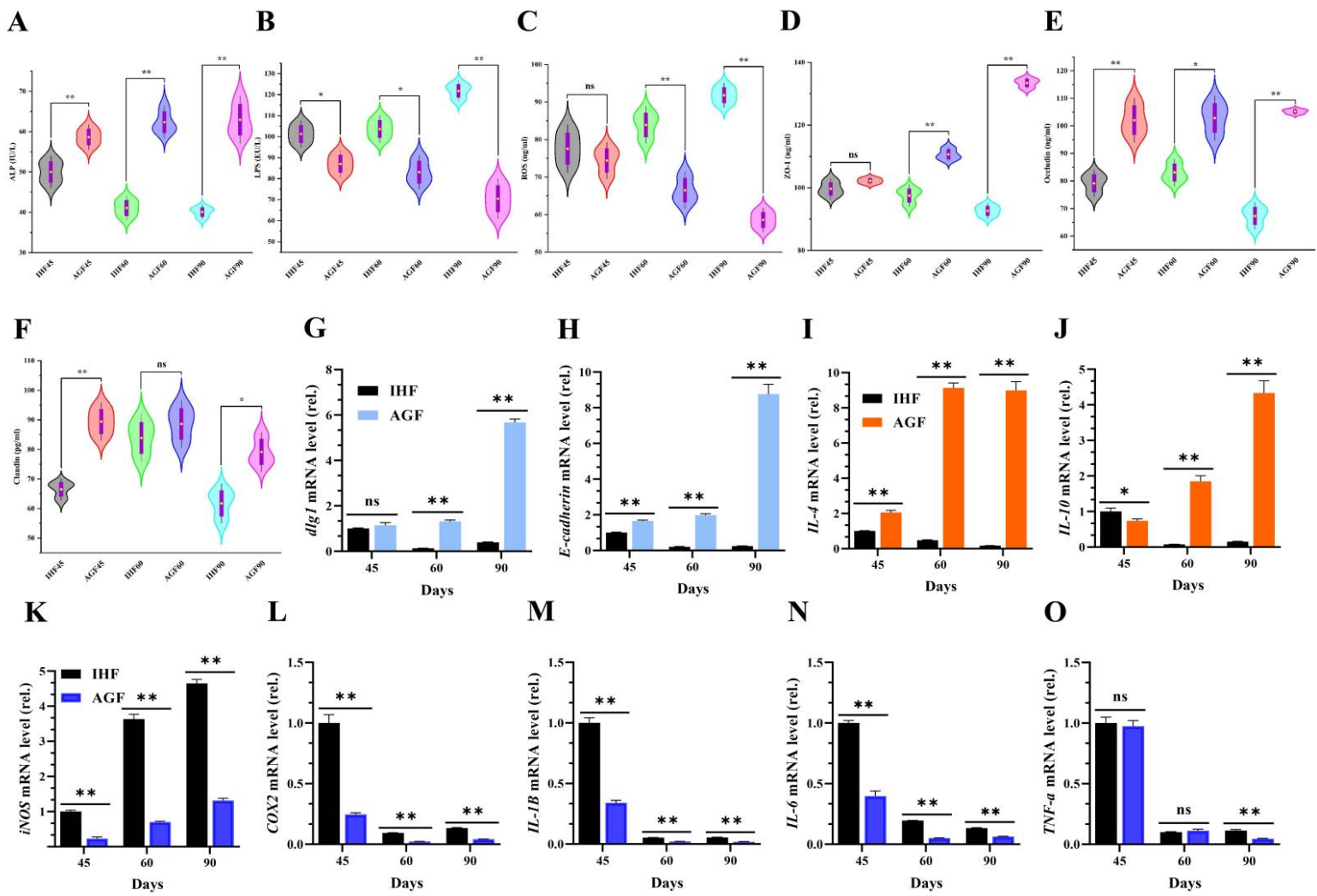

**Figure. 7**

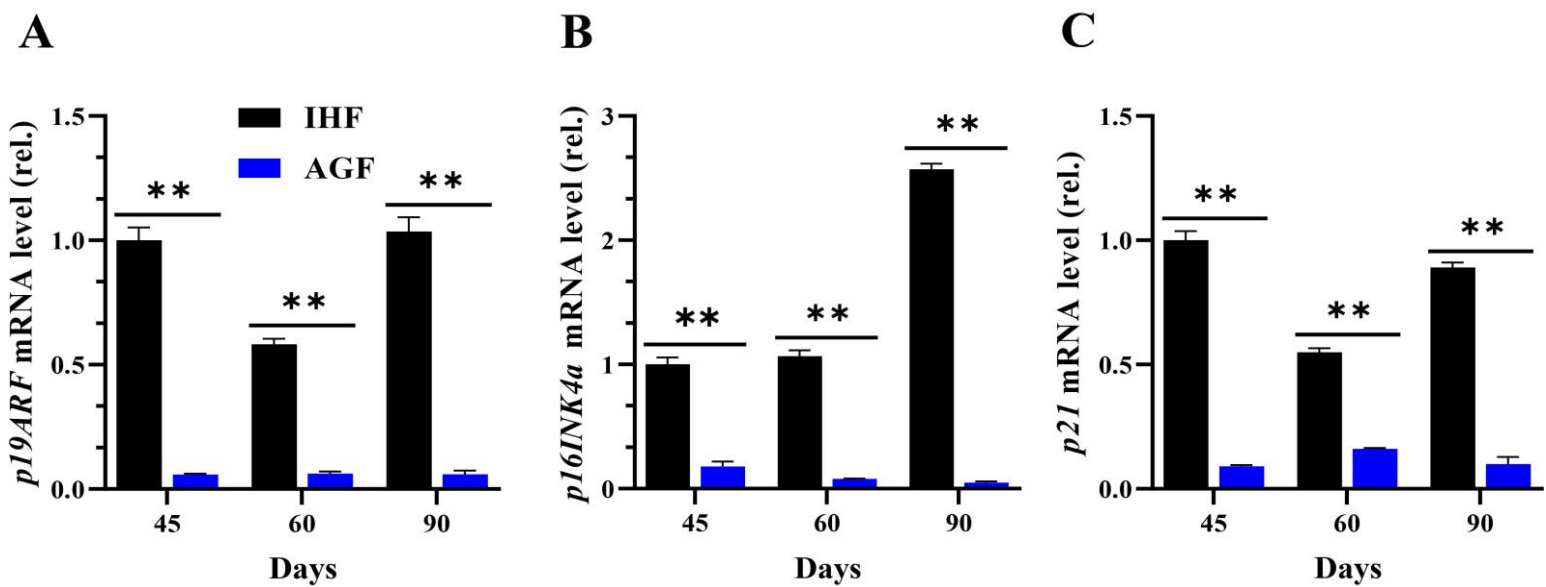

**Figure. 8**

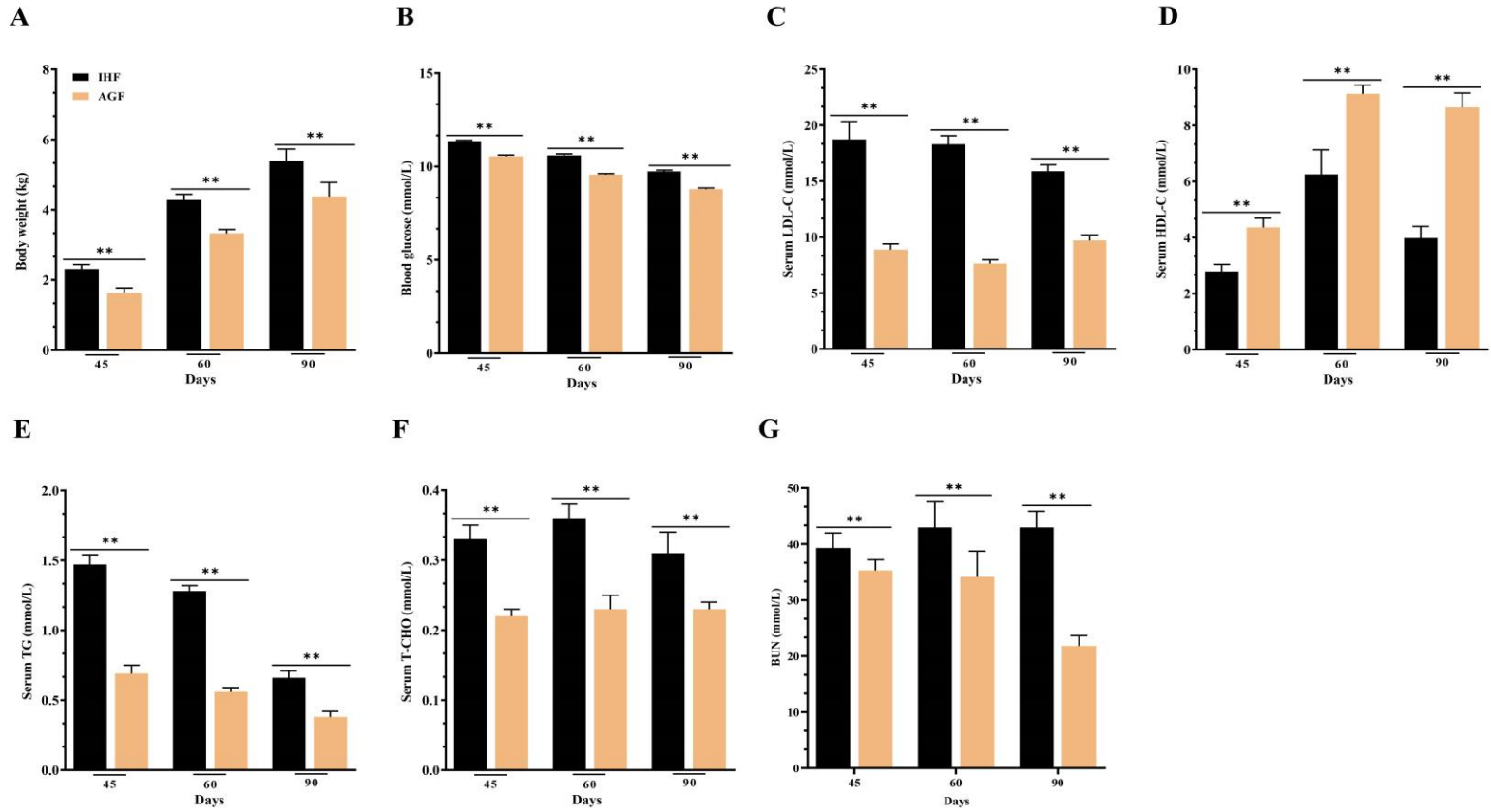

# Figure. 9

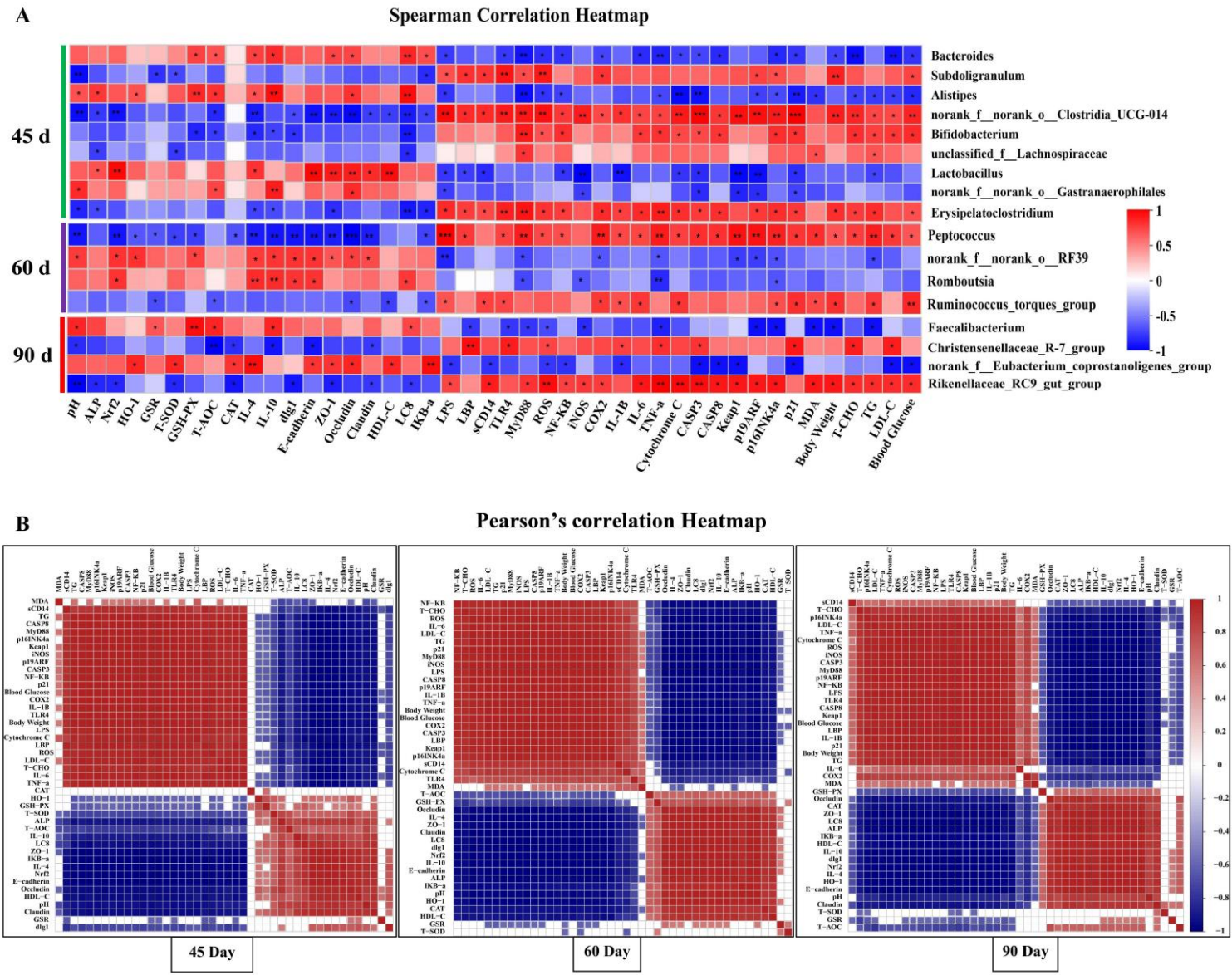

**Figure. 10**

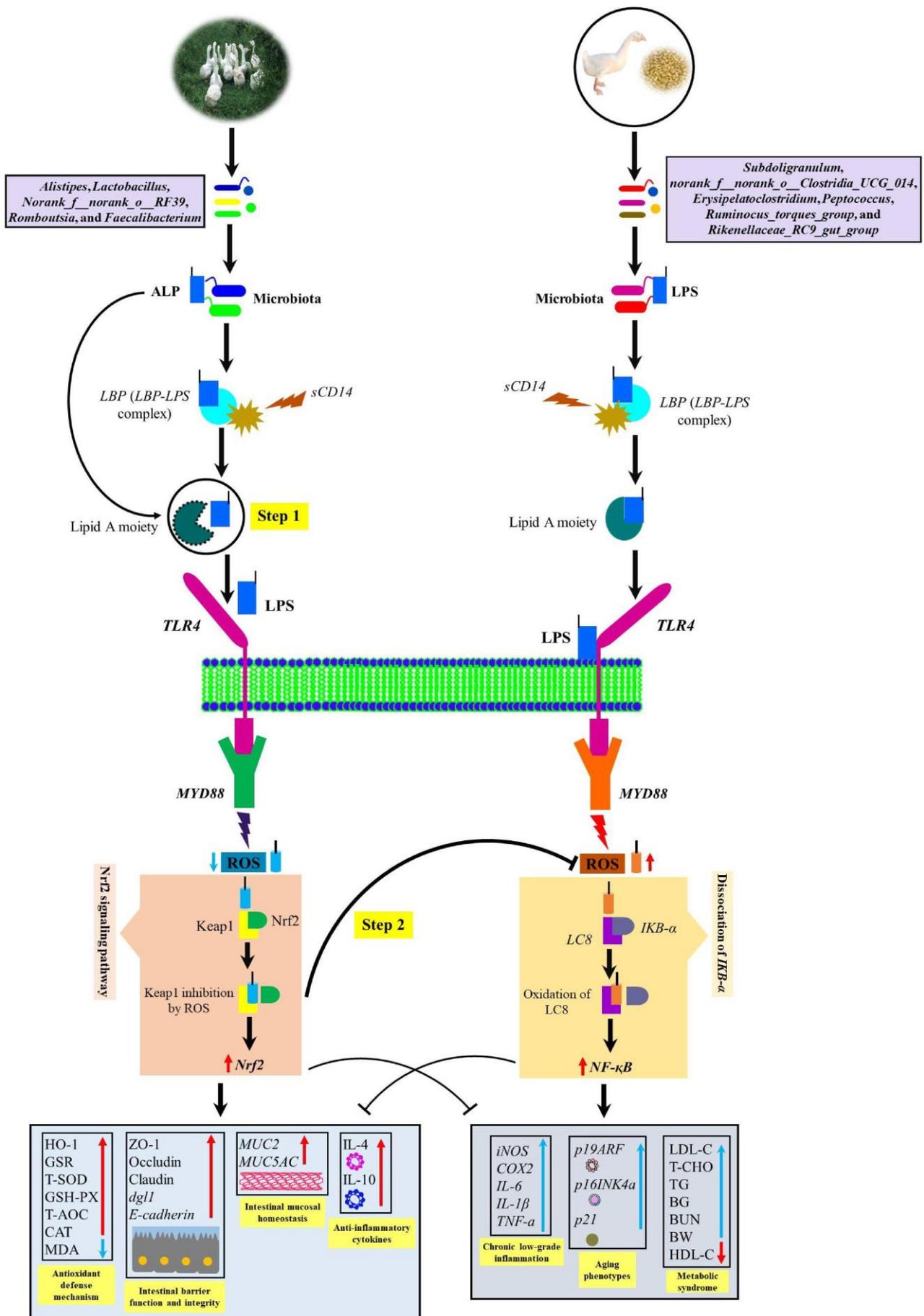

# Supplementary figures

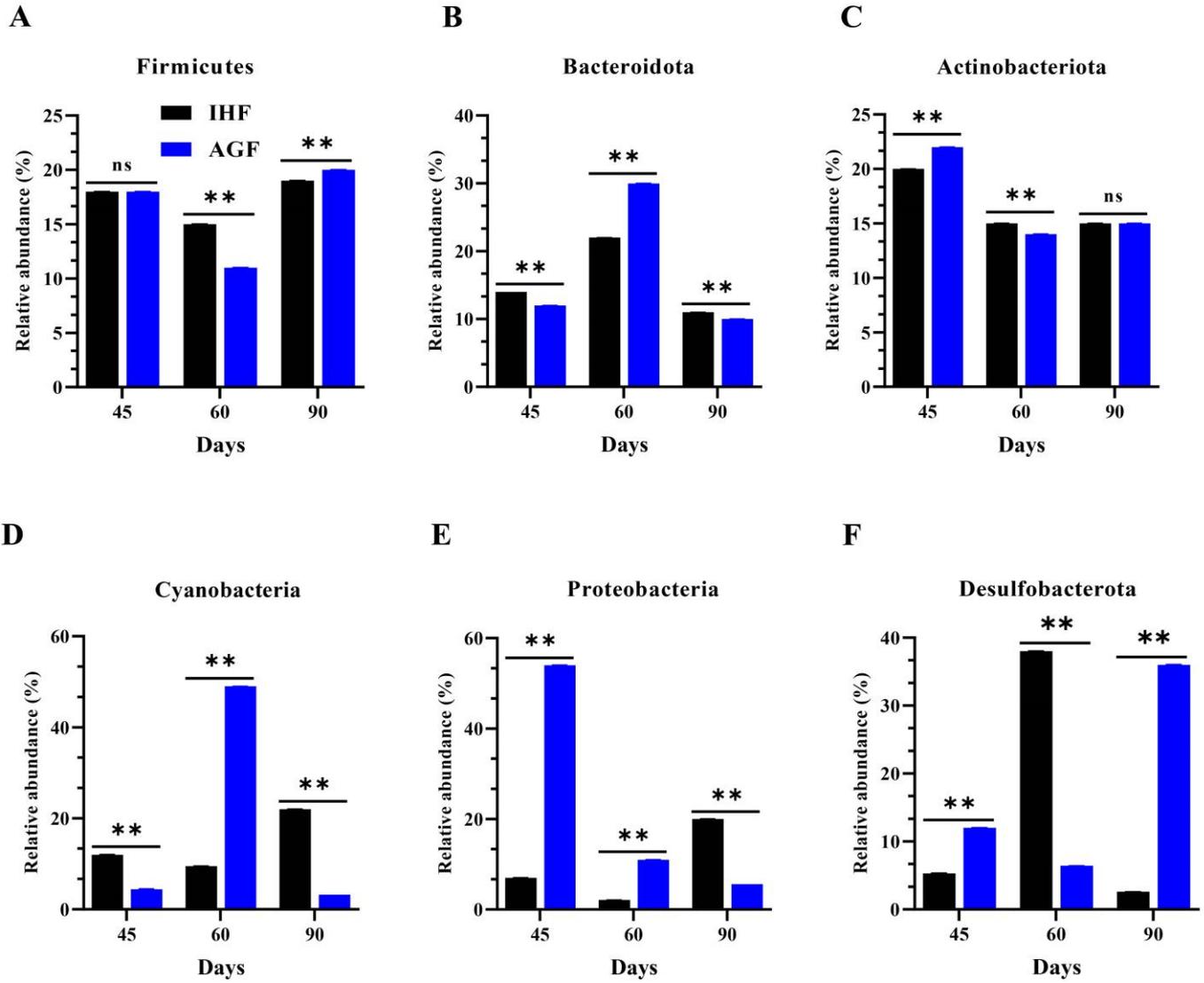

**S2-1_FIG.**

A

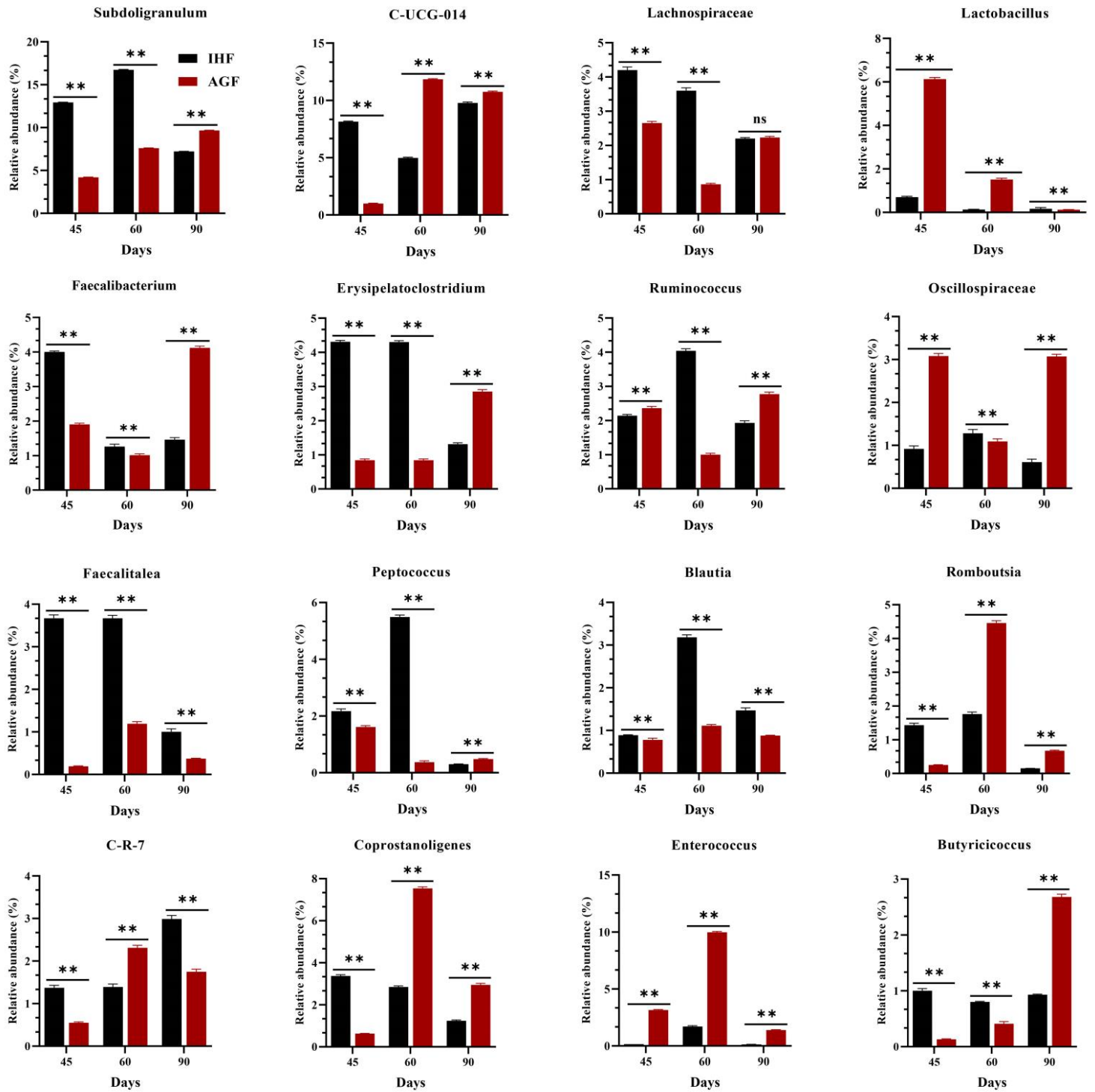

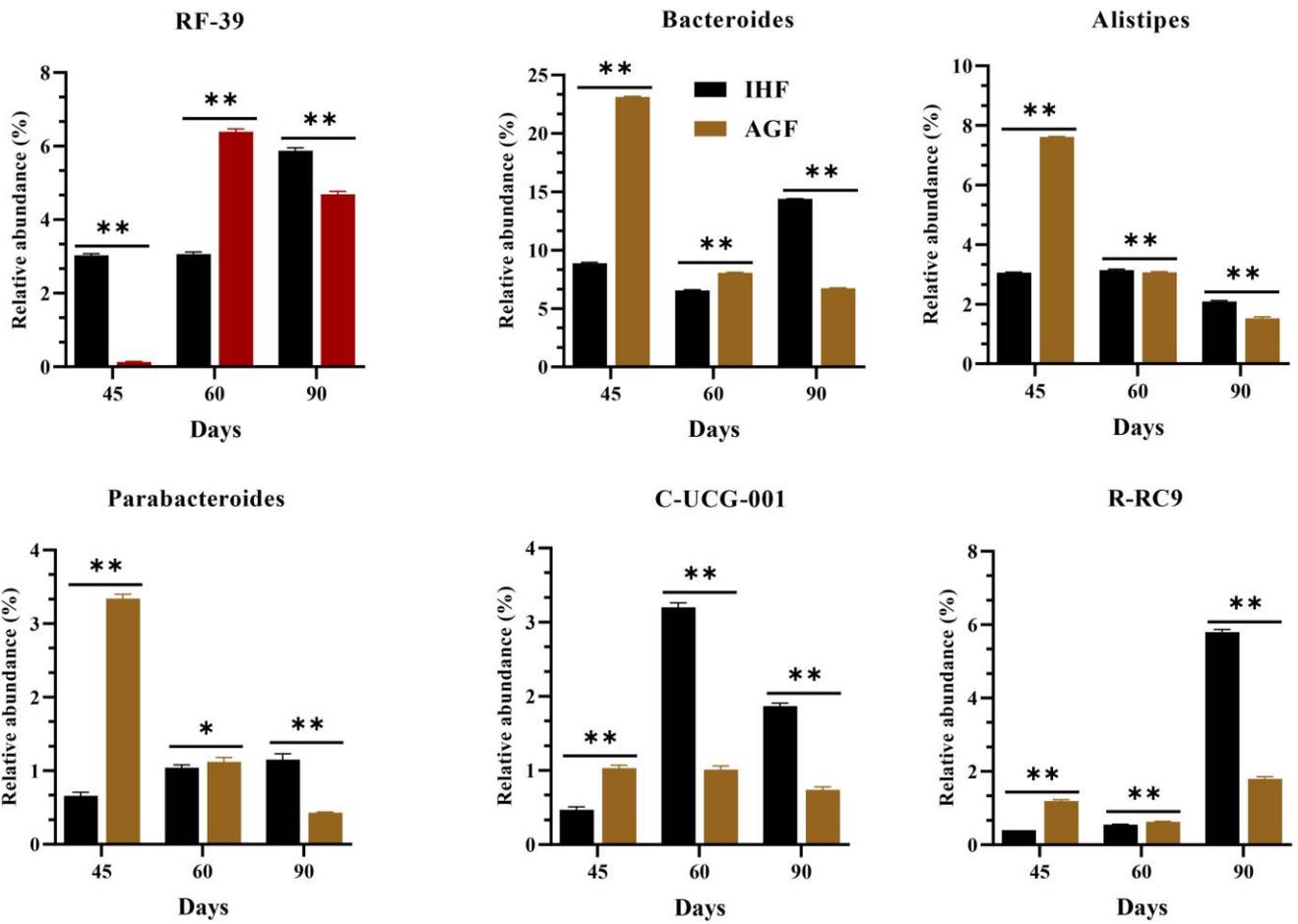

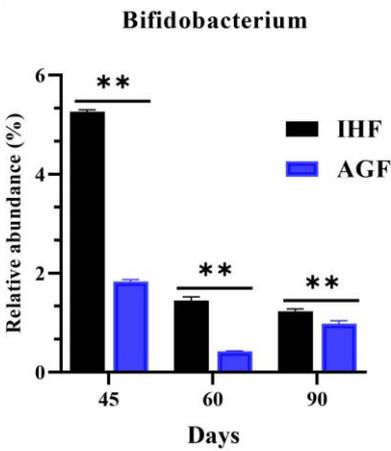

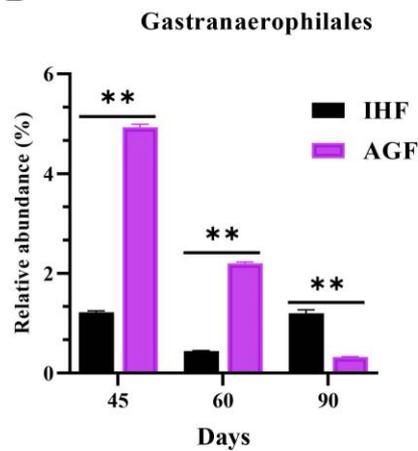

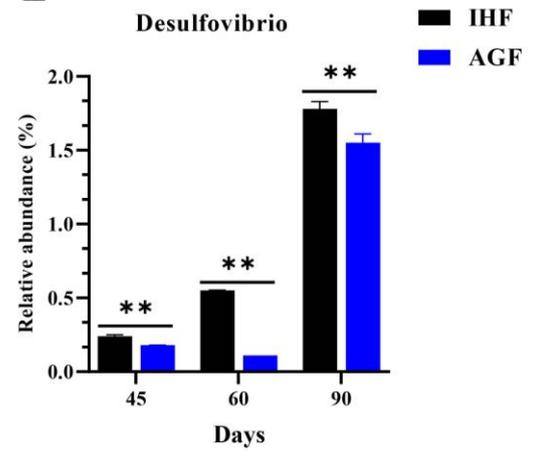

**S3_FIG.**

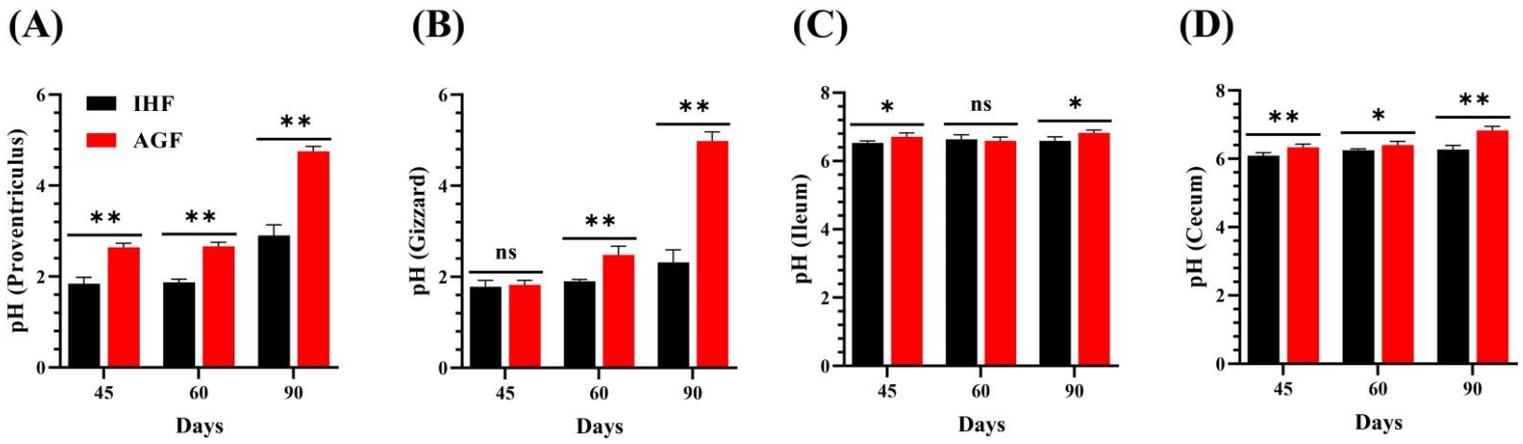

**(A)** **(B)** **(C)** **(D)**

**S4_FIG.**

**(A)**

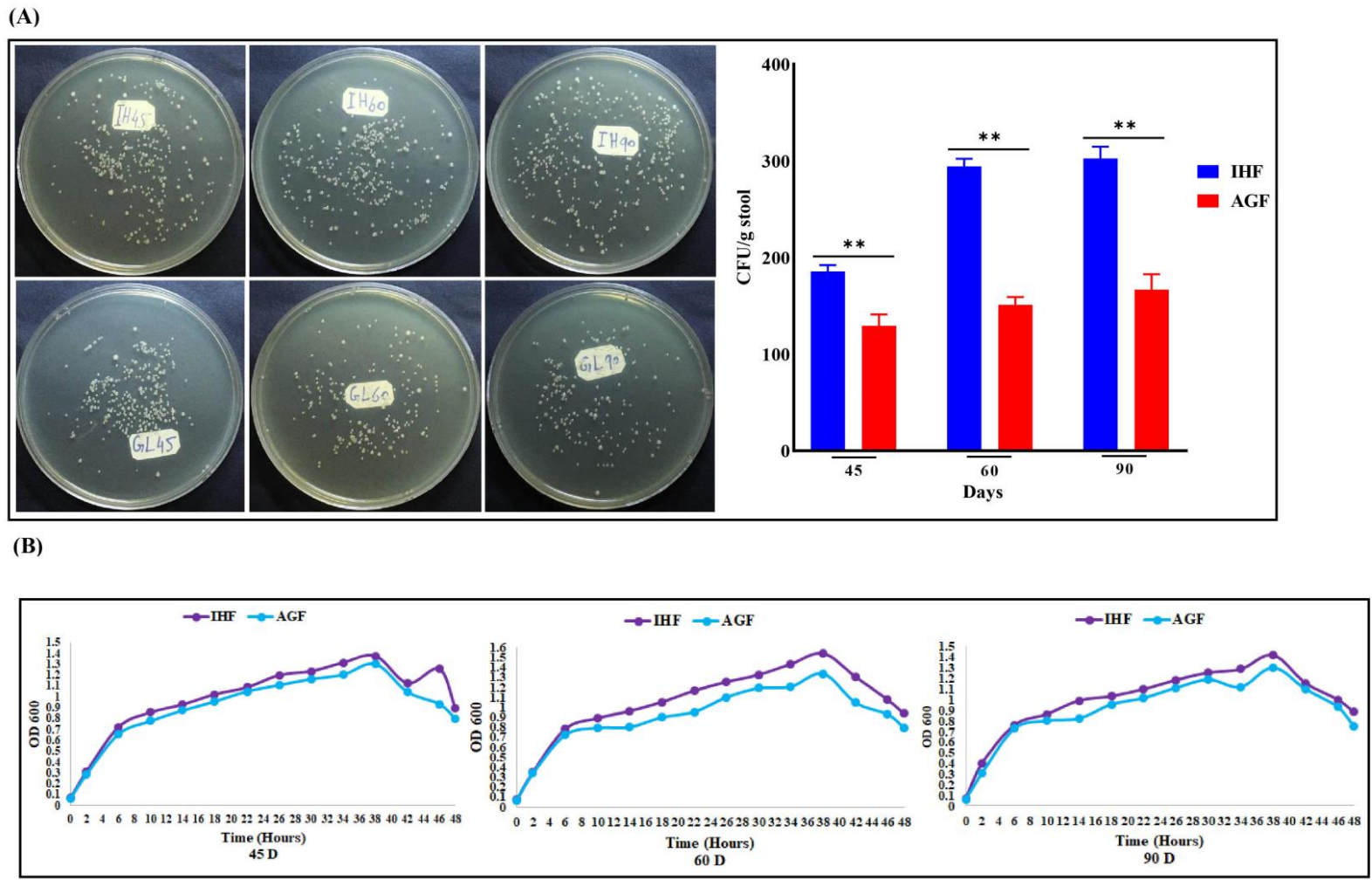

**(B)**

**S5_FIG.**

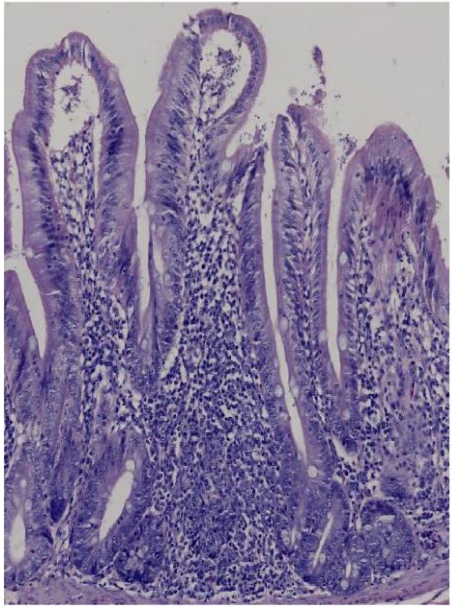
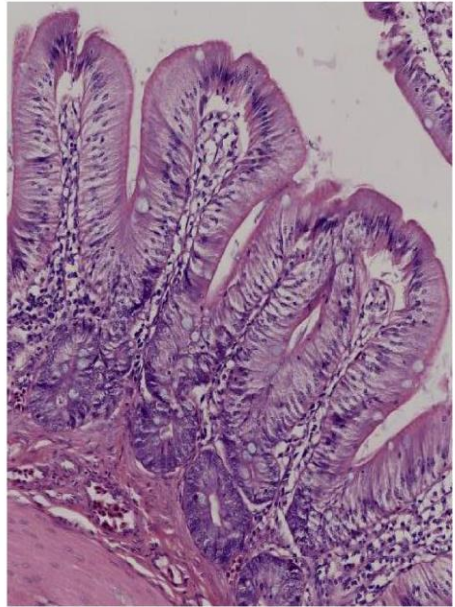
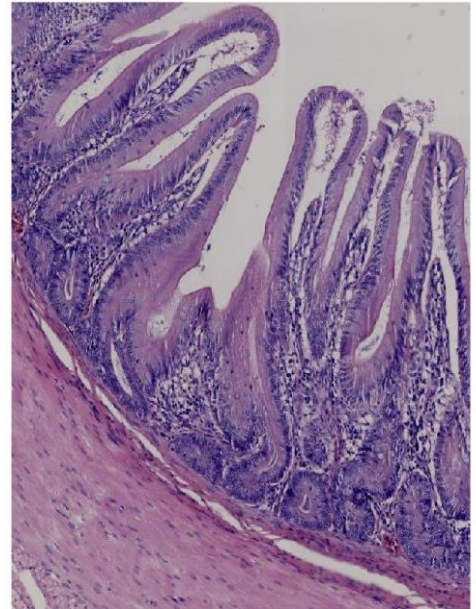

IHF

**45 D**          **60 D**          **90 D**

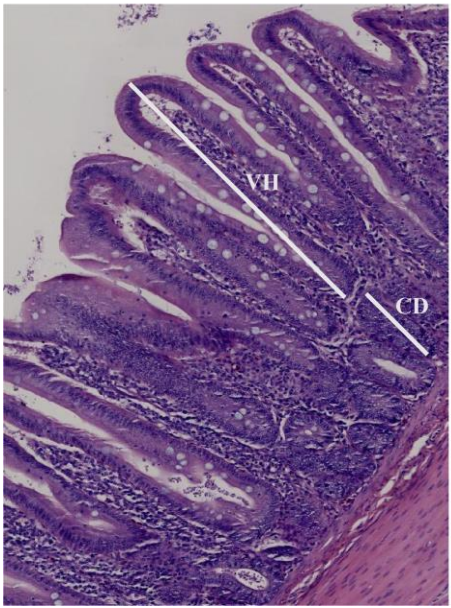
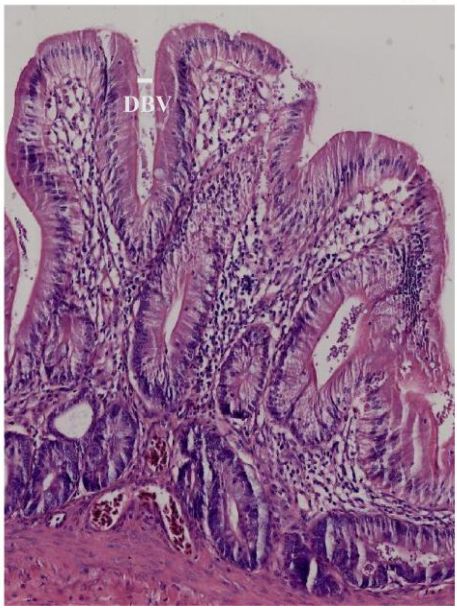
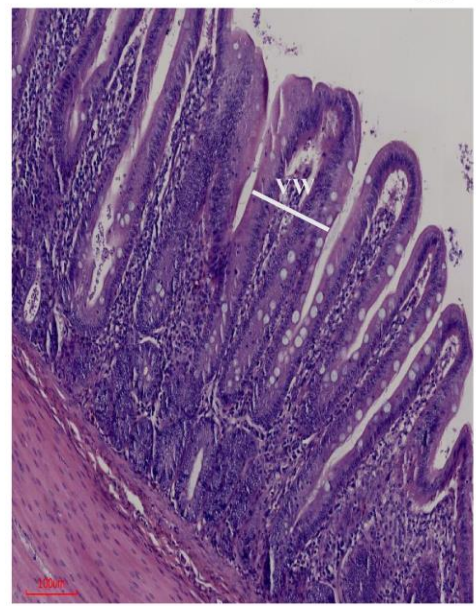

AGF

**S6_FIG.**

**A**

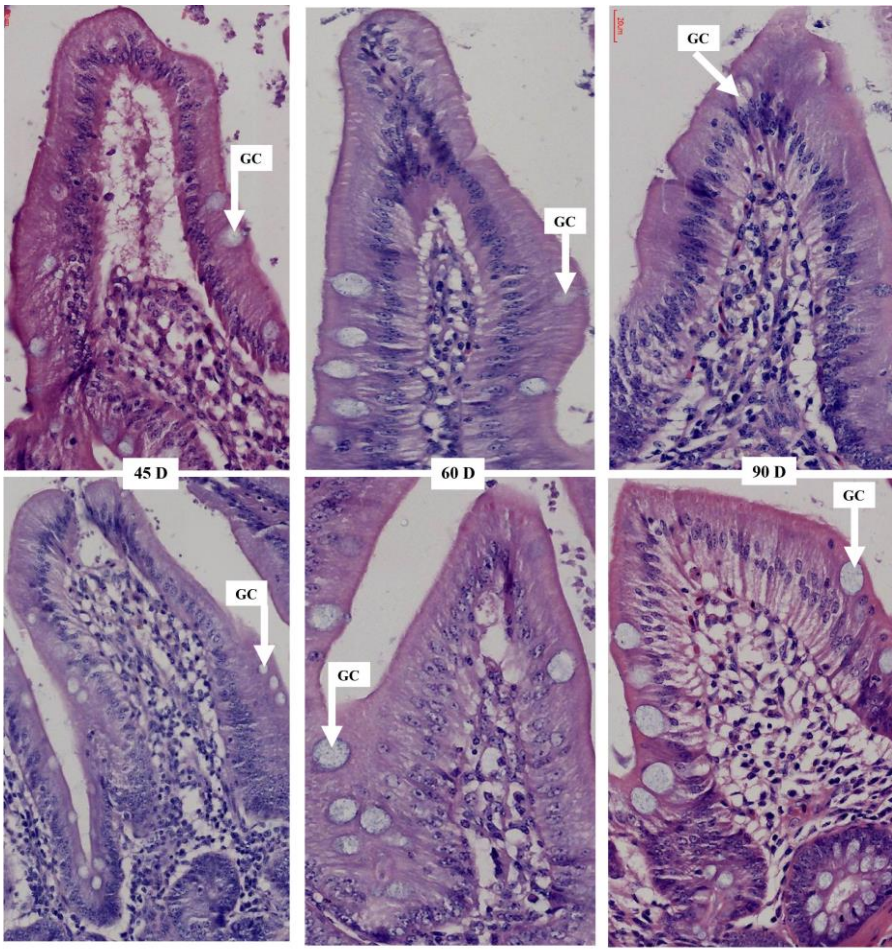

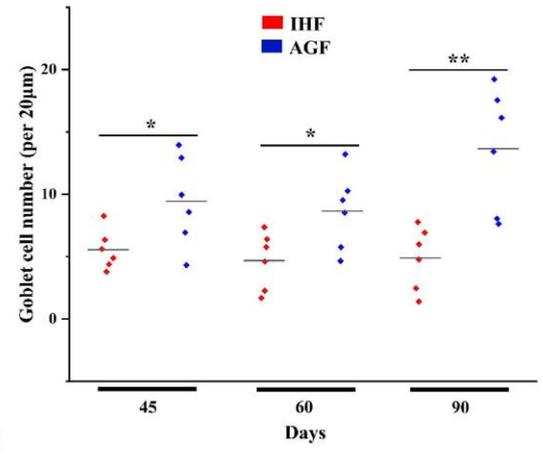

**S7_FIG.**

A

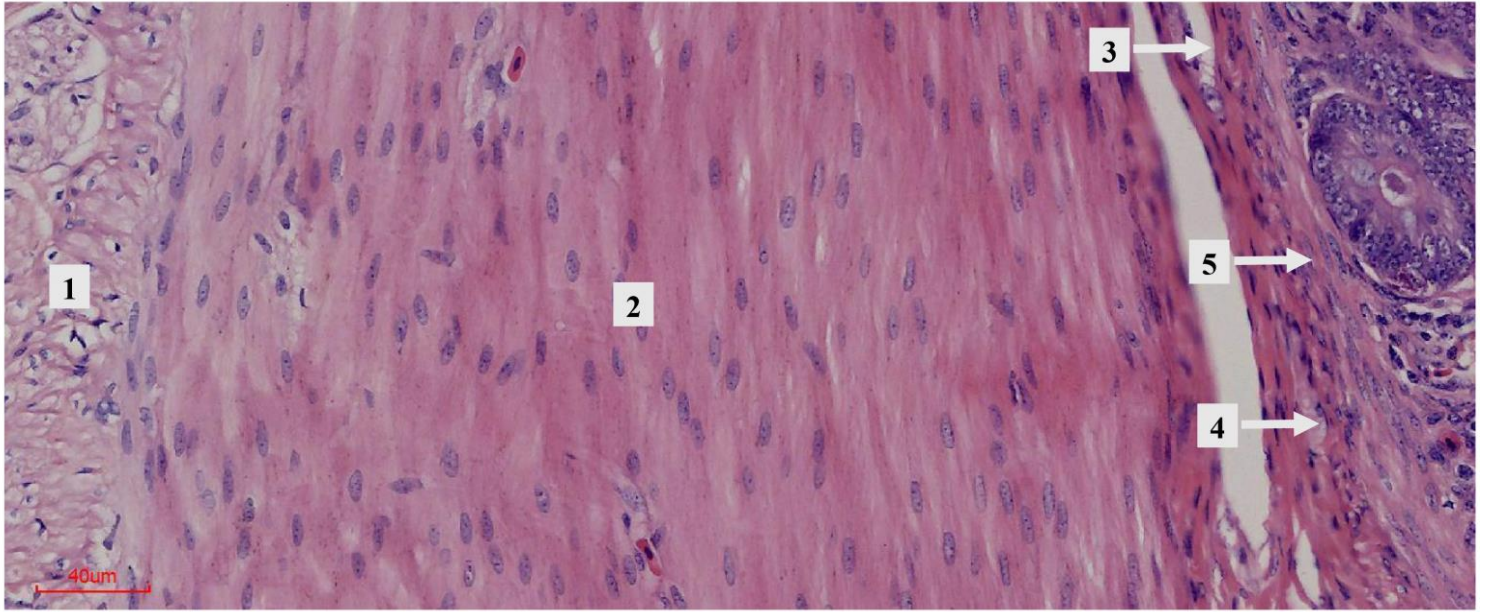

B

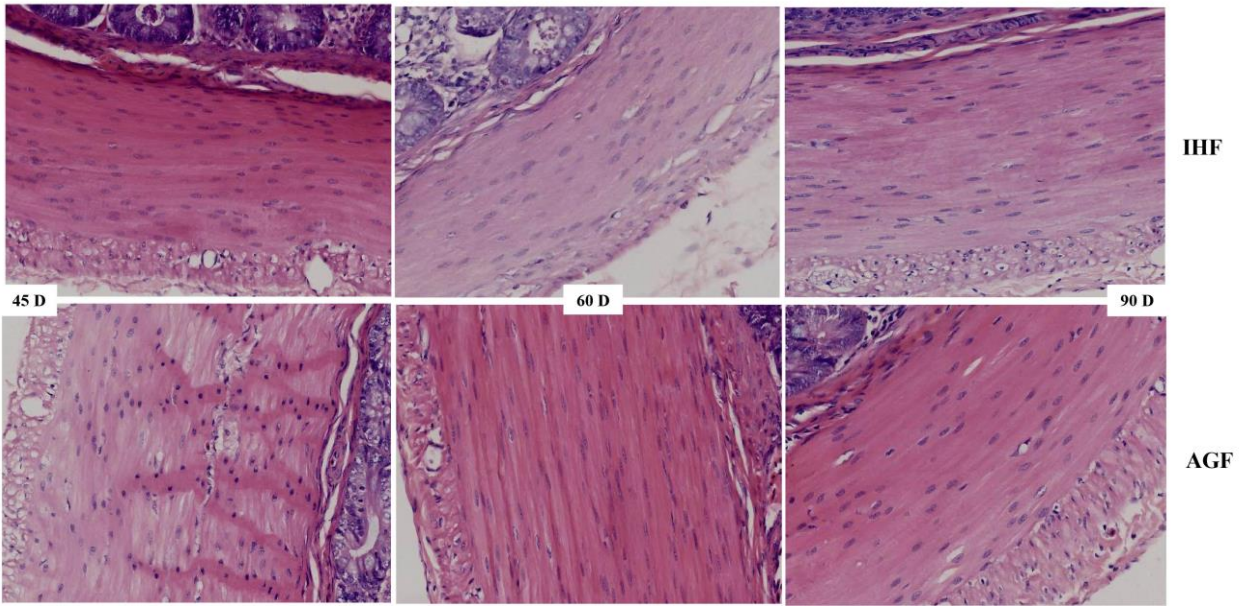

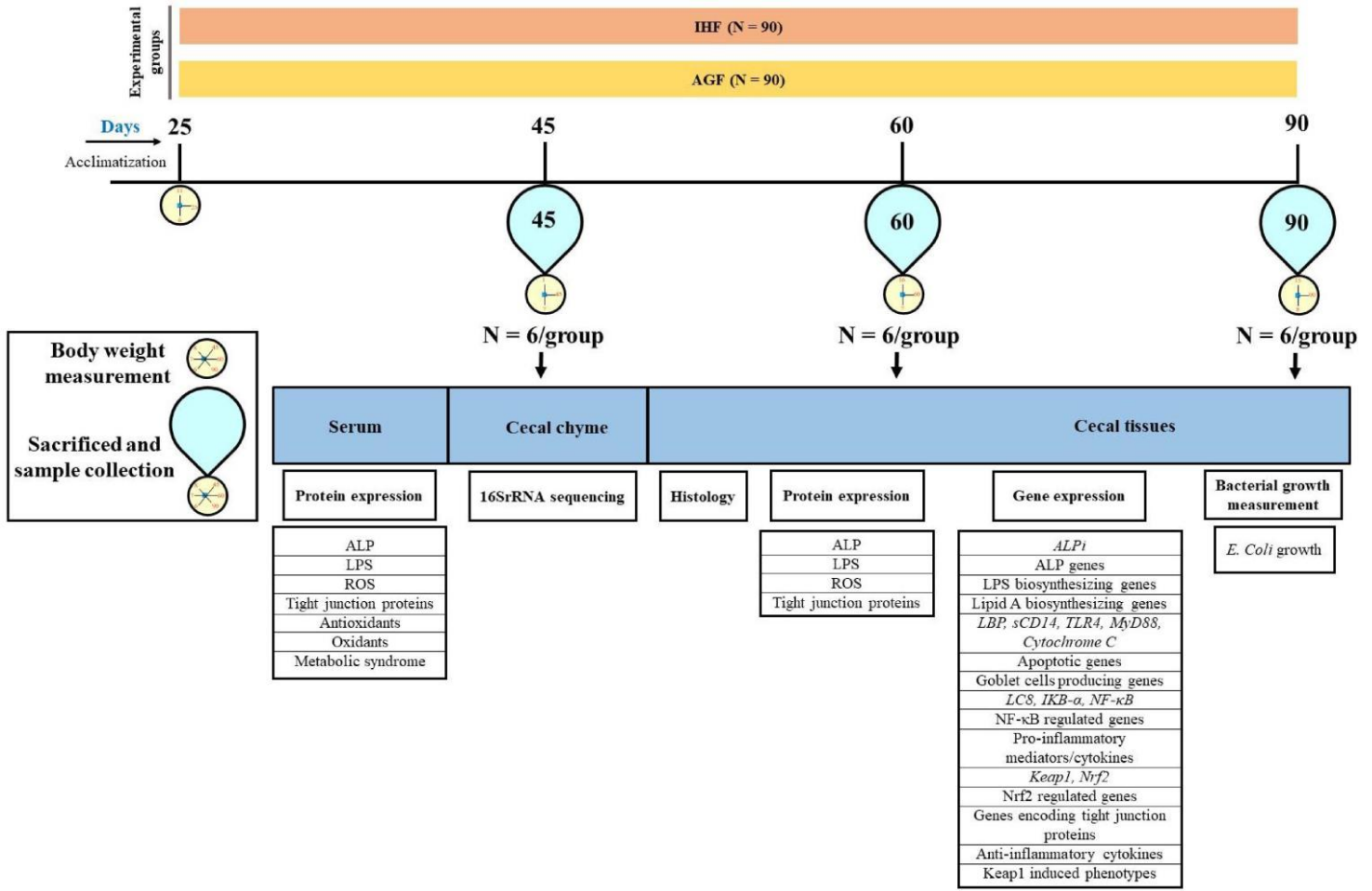

# Supplementary tables

## Supplemental Table 1. Primer sequences used for quantitative real-time PCR

| Gene | Forward primer (Sequence 5′- 3′) | Reverse primer (Sequence 5′- 3′) |
|------|----------------------------------|----------------------------------|
| IKB-α | AAAGCTGGATGTGACCTGGA | GTTGTAGTTGGTTGCCTGCA |
| NF-KB | GGCAGAGATGGTGGAAGACT | GTTTGCCATCACCACCATGT |
| NRF2 | CGCCTTGAAGCTCATCTCAC | CCTCTCCTGCGTATATCCCG |
| iNOS | CTCATTCTCCAAGCGAACGG | GCACTCCTATCTCTGTCCCC |
| COX2 | GGTTCTACAATGGAGAGCGC | TGTTCTTGCCACTTGAGCTG |
| IL1-B | CACATCACAACCCACAGCAA | CTGCCCCTTCCGTCTTCTTA |
| IL-6 | GGAAGACCCTTGCTCTCCTT | TGGAGCCAGAAGATGAGTGG |
| TNF-α | GGTCCACAACGAGTTCATCC | AGGAGGAGGAGGAGATGGAG |
| β-actin | CAACGAGCGGTTCAGGTGT | TGGAGTTGAAGGTGGTCTCG |
| Keap1 | CTGAACGAGGCCCTCAAGTA | CAAACTCGTAGCGGGGAATG |
| MUC2 | GCCTTTCACTCAGCAGCTTT | CCATGACTCACCTGGCTGTA |
| MUC5AC | GGAAAACACGGGCTGATTGT | CTCTGACGTAGCTGGAGAGG |
| CASP3 | TCAGAGGTGACAAGTGCAGA | TCAGCACCCTACACAGAGAC |
| CASP8 | CACCACTGGCATCACAAACA | ATTCCTCCTGTCAGTCCAGC |
| NQ01 | GACCCCAAGCACTTCGTCTA | CGAAGCCTTGGATGATGACG |
| GCLC | GGAGAGTGGAATTCAGGCCT | AGAGCCACATCCATCCACAA |
| GCLM | TGCCCCACCTCCTATTGAAG | TGTGAGATCAGGTGGCATCA |
| GSTA4 | GCACCTCTGTTCCAAAGCAA | AATCCAGGACACAGCCATCA |
| IL-8 | TTCTCCTGATTTCCGTGGCT | TTCAACGTTCTTGCAGTGGG |
| CCL2 | ACTGGGAGATGTGGATGTGG | CCATACAGCTCAGCAAACCC |
| BIRC3 | TGGCCCCTGATGTTTCTCTT | AGTGTCTCCGATGCTCTGAC |
| PLAU | ACCCCTCAAAAGCTACCGAA | TCCCATGCTTCTGTTTGTGC |
| P21 | CCACGACCAGCTCCAGAAT | CTTGCCAAGACTGAGGACCT |
| P19ARF | GCTCTCAAAGGACAGCGAAC | AGGATCTGGAAGGAGCTGTG |
| TLR4 | AGGGCTACAGGTCAACAGAC | GACGTTCACCAGCCGAATAC |
| MYD88 | CTGCGTCTTTGATCGGGATG | GGCTTTGCACTTCACTGGAA |
| p16INK4α | GCCGAGCACCAGAATTTGAT | CTGCTGGGTTTTGAGGGAAC |
| lpxA | TGATGCTCCTCAGACTGCTC | GCCAACAGAGTGCTCCAAAT |
| lpxB | GAGACTTCTTCATCCGCTGC | TCCACCTGAGCAATGTCTGT |
| lpxC | GTTGTTGCAAGGAGGGTGAG | TCCAGCCTCTCACCAAACAT |
| lpxD | TCCGTTGGTCTGTCTGCTTA | GCCCTGTCTAGTGCTGAAGA |
| rfaK | TGTTCACTTCTCCTAGACAGCA | CTTGTCCTTTCAGAAGTGGCA |
| rfaL | CCTCTTCTGCTGGCACTACT | TTCCCATCCCCAGCAGAAAT |
| LBP | CTGTCTGTGGCTCTGATGGA | AGCACCGAACCAGTCCTTTA |
| sCD14 | TGGATGGGGCTGTGGTATAC | GTCACAGGTCAGCTCTCGAT |
| ALPi | CCTAGGGATGAAGCAGGGAG | TGGGCCACTGTCATCTCATT |
| CG5150 | ACGAGCAGAACTACATCCCC | TAGCAGAGGAGGAGGAGGAG |
| CG10827 | GAGGGCAAAGAATGGGCAAA | ACCAAGAAAACAGCAGCCAG |
| E-cadherin | GGTGACGGTGGAGAACAAAG | CTGGGCTGTGTAGGATGTGA |
| dlg1 | AGTGATATTGACCGTGGCCA | AATCACGTCCATCCACCTCA |
| IL-4 | TGACAGGGTATTGGTCCACC | ACGGAAGAAGCAGAAGGTGA |
| IL-10 | CATCAAGAACAGCGAGCACC | GCACCCACCTTTTCAAACGT |
| LC8 | GAGGAGATGCAGCAGGACTC | GAGTCACGTAGCTGCCAAAG |
| Cytochrome C | GGGTTACATAGGGCGGAGTC | GCCTCCCTTCTCAACCGTAT |

**Supplemental Table 2. Effect of artificial pasture grazing system on pH of proventriculus, gizzard, ileum, and cecum of meat geese.** Data expressed as mean ± SEM

| Age, d | Parameters | IHF | AGF | P-value |
|--------|------------|-----|-----|---------|
| 45 d | | 1.84±0.14 | 2.64±0.09 | 2.3704E-05 |
| 60 d | Proventriculus | 1.87±0.08 | 2.66±0.1 | 7.28674E-06 |
| 90 d | | 2.9±0.24 | 4.75±0.11 | 1.10084E-05 |
| 45 d | | 1.78±0.14 | 1.82±0.11 | 0.330742017 |
| 60 d | Gizzard | 1.9±0.04 | 2.48±0.21 | 0.001 |
| 90 d | | 2.32±0.27 | 4.98±0.2 | 1.58863E-09 |
| 45 d | | 6.53±0.06 | 6.71±0.11 | 0.023 |
| 60 d | Ileum | 6.64±0.13 | 6.59±0.10 | 0.41 |
| 90 d | | 6.59±0.11 | 6.82±0.09 | 0.02 |
| 45 d | | 6.09±0.1 | 6.34±0.1 | 0.001 |
| 60 d | Cecum | 6.25±0.04 | 6.4±0.12 | 0.01 |
| 90 d | | 6.27±0.13 | 6.83±0.13 | 1.01066E-05 |

**Supplemental Table 3. Effects of artificial pasture grazing system on the cecal morphology of meat geese.** Data expressed as mean ± SEM.

| Parameters | 45 D | | 60 D | | 90 D | | P-values | | |
|------------|------|------|------|------|------|------|------|------|------|
| | IHF | AGF | IHF | AGF | IHF | AGF | 45 D | 60 D | 90 D |
| Villus height (um) | 63.69±10.22 | 96.89±16.96 | 62.19±12.74 | 112.96±21.23 | 68.93±12.64 | 93.71±17.14 | <0.001 | <0.0003 | <0.01 |
| Villus width (um) | 32.97±3.69 | 41.84±9.86 | 25.97±7.48 | 37.31±9.56 | 23.37±6.59 | 41.54±6.6 | <0.033 | <0.023 | <0.0004 |
| Surface area (um²) | 4379.64±969.12 | 11776.67±5400.16 | 3762.17±1454.98 | 7901.81±2201.51 | 3960.01±596.81 | 10442.38±2385.9 | <0.004 | <0.0016 | <0.0004 |
| Crypt depth (um) | 24.82±9.32 | 15.5±1.61 | 24.12±5.57 | 16.33±6 | 25.4±5.42 | 17.37±3.42 | <0.018 | <0.02 | <0.01 |
| Villus height/Crypt depth | 2.13±0.17 | 1.86±0.73 | 1.64±0.37 | 1.56±0.35 | 1.38±0.4 | 1.72±0.54 | 0.2 | 0.37 | 0.1 |
| Distance between villi (um) | 3.37±0.42 | 6.02±0.88 | 3.12±0.44 | 6.46±0.59 | 3.51±0.99 | 6.51±1.16 | <0.00003 | <3E-07 | <0.005 |

**Supplemental Table 4. Effect of different feeding systems on the numbers of apoptotic cells per field in the villus of cecal tissues.** In-house feeding system (IHF) and artificial pasture grazing system (AGF). Data expressed as mean ± SEM.

| Age, d | IHF | AGF | P-value |
|---|---|---|---|
| 45 | 20.8±8.73 | 7.58±2.64 | 0.003 |
| 60 | 14.25±4.88 | 7.11±1.6 | 0.01 |
| 90 | 10.84±2.32 | 6.93±0.76 | 0.001 |

**Supplemental Table 5. Effect of different feeding systems on the thickness of ileal muscular tonic and muscularis mucosa (50μm).** In-house feeding system (IHF) and artificial pasture grazing system (AGF). Data expressed as mean ± SEM.

| Parameters | 45 D | | 60 D | | 90 D | | P-value | | |
|---|---|---|---|---|---|---|---|---|---|
| | IHF | AGF | IHF | AGF | IHF | AGF | 45 D | 60 D | 90 D |
| Inner layer (um) | 48.81±8.64 | 61.39±6.06 | 47.42±21.93 | 80.35±8.67 | 39.79±17.75 | 68.34±13.34 | <0.00765 | <0.003 | <0.005 |
| Outer layer (um) | 8.54±2 | 10.88±1.41 | 7.84±1.69 | 17.25±1.08 | 8.49±2.18 | 20.26±4.31 | <0.02 | <2.18E-07 | <6.90E-05 |
| Total (um) | 57.35±10.2 | 72.26±5.37 | 55.26±22.19 | 97.6±9.02 | 48.27±17.3 | 88.6±14.27 | <0.005 | <0.001 | <0.001 |
| Relative thickness of muscular tonic (um) | 24.84±4.51 | 44.49±3.26 | 12.88±4.84 | 29.33±2.7 | 9.07±3.46 | 20.44±4.37 | <2.94E-06 | <1.34E-05 | <0.0003 |
| Inner layer (um) | 2.71±0.32 | 3.87±0.71 | 3.13±1.27 | 7.75±2.55 | 3.1±0.88 | 4.98±1.47 | <0.002 | <0.001 | <0.01 |
| Outer layer (um) | 1.08±0.28 | 1.59±0.25 | 1.27±0.28 | 1.53±0.19 | 1.06±0.28 | 1.48±0.22 | <0.004 | <0.04 | <0.01 |
| Total (um) | 3.78±0.36 | 5.47±0.59 | 4.39±1.16 | 9.28±2.52 | 4.16±0.93 | 6.46±1.51 | <0.0001 | <0.001 | <0.005 |
| Relative thickness of muscularis mucosa (um) | 1.64±0.15 | 3.39±0.55 | 1.03±0.31 | 2.79±0.77 | 0.78±0.21 | 1.5±0.45 | <0.00001 | <0.0002 | <0.003 |